\documentclass[pre,twocolumn,superscriptaddress,citeautoscript]{revtex4-1}

\usepackage{epsfig}
\usepackage[usenames, dvipsnames]{color}
\usepackage{natbib}
\usepackage{xifthen}
\usepackage{bm}
\usepackage[version=3]{mhchem} 
\usepackage{mciteplus}
\usepackage{multirow}
\usepackage{graphicx}
\usepackage{dcolumn}
\usepackage{bm}
\usepackage{mathptmx,amsmath,amssymb,txfonts}
\usepackage{subfig}
\usepackage{blindtext}

\begin{document}

\title{The Gregory-Newton Problem for Kissing Sticky Spheres}

\author{Lukas Trombach}
\affiliation{Centre for Theoretical Chemistry and Physics, the New Zealand Institute for Advanced Study, Massey University Auckland, Private Bag 102904, 0632 Auckland, New Zealand}
\author{Peter Schwerdtfeger} 
\email[Email: ]{p.a.schwerdtfeger@massey.ac.nz}
\affiliation{Centre for Theoretical Chemistry and Physics, the New Zealand Institute for Advanced Study, Massey University Auckland, Private Bag 102904, 0632 Auckland, New Zealand}
\affiliation{Centre for Advanced Study (CAS) at the Norwegian Academy of Science and Letters, Drammensveien 78, NO-0271 Oslo, Norway}


\begin{abstract}
All possible non-isomorphic arrangements of 12 spheres kissing a central sphere
    (the Gregory-Newton problem) are obtained for the sticky-hard-sphere (SHS) 
    model, and subsequently projected by geometry optimization onto a set of
    structures derived from an attractive Lennard-Jones (LJ) type of potential.  
    It is shown that all 737 derived SHS contact graphs corresponding to the 12 outer
    spheres are (edge-induced) subgraphs of the icosahedral graph. The most
    widely used LJ(6,12) potential has only one minimum structure corresponding
    to the ideal icosahedron where the 12 outer spheres do not touch each
    other. The point of symmetry breaking away from the icosahedral symmetry
    towards the SHS limit is obtained for general LJ($a,b$)
    potentials with exponents $a,b\in \mathbb{R}_+$. Only if
    the potential becomes very repulsive in the short-range, determined by the
    LJ hard-sphere radius $\sigma$, symmetry broken solutions are observed.
\end{abstract}
\maketitle

\section{Introduction}

The arrangement of $N$ points on the surface of a sphere corresponding to the placement of $N$ identical non-overlapping spheres around a central sphere is called a spherical packing. To achieve optimal packings for spheres is known as the Tammes problem, originally posed in 1930 to study the distribution of pores on pollen grains \cite{tammes_1930}: It is to determine the largest diameter and distribution that $N$ equal non-overlapping spheres may have when packed onto the surface of a sphere of radius 1 (unit sphere). Alternatively, if the centre of each sphere is considered as the vertex of a polyhedron, the graph theoretical problem is to find the polyhedron that maximizes the shortest edge lengths with fixed distance to the central vertex. The Tammes problem has been solved exactly for $3 \le N \le 14$ and $N = 24$ \cite{Robinson_1961,Musin_2015}.

\begin{figure}[htb]
    \centering
    \includegraphics[width=.4\textwidth]{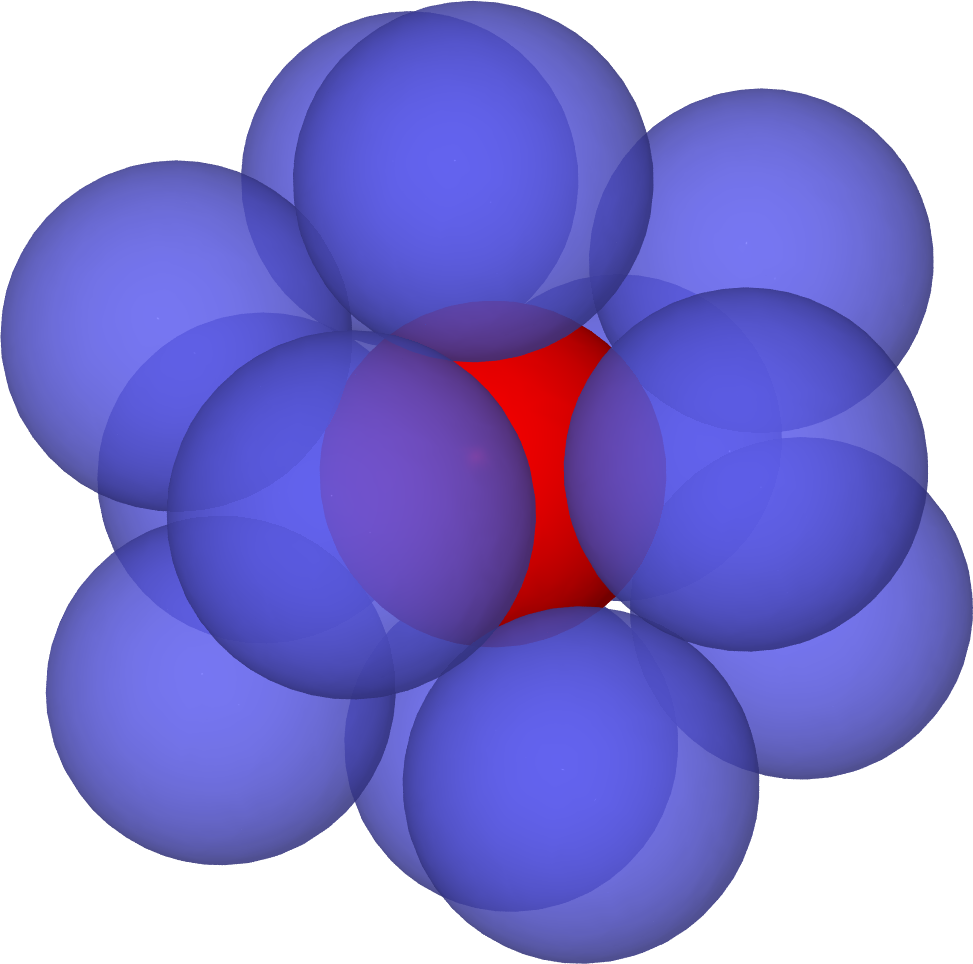}
    \includegraphics[width=.45\textwidth]{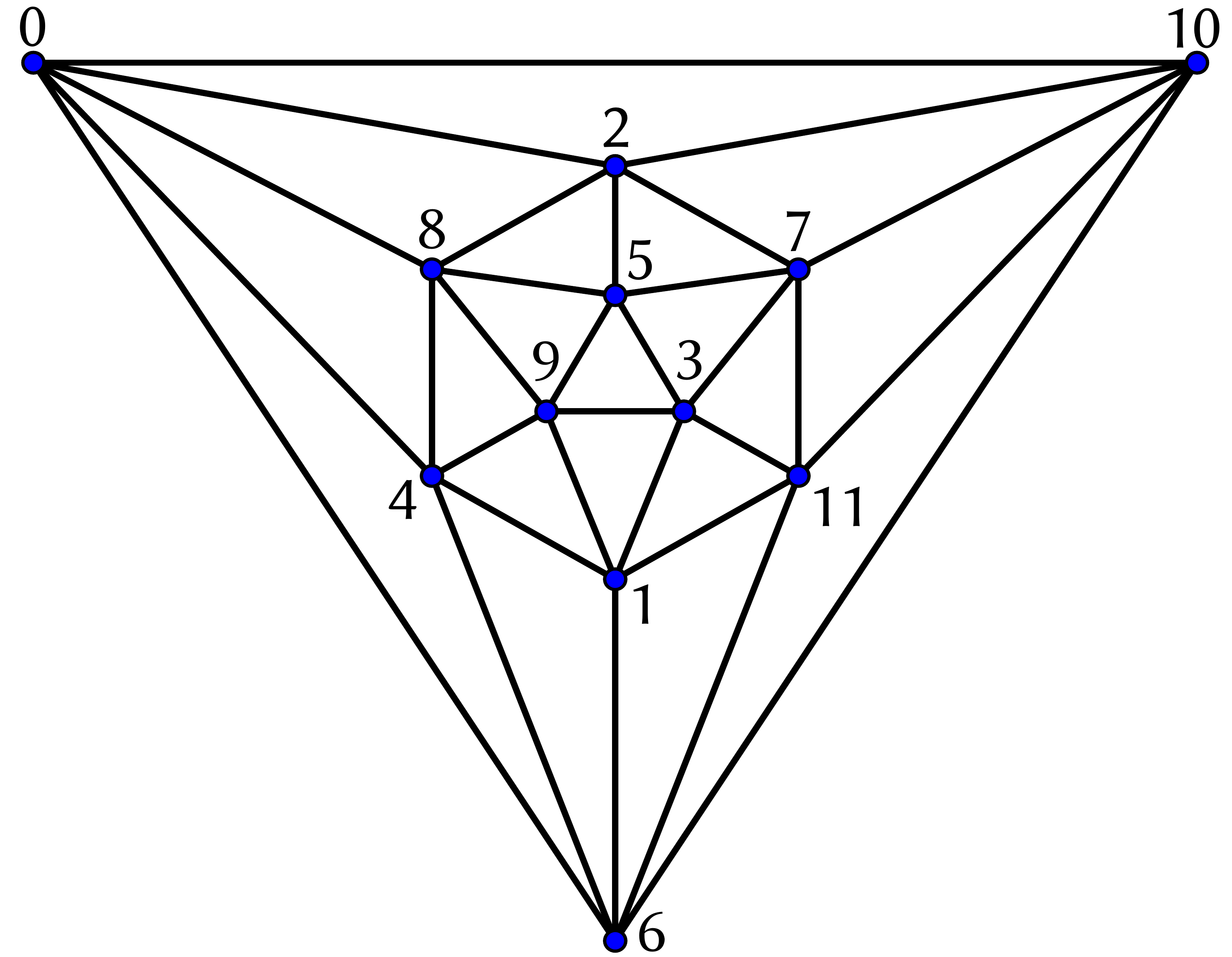}
    \caption{Top: Symmetric realization of $N_k(3)=12$ for unit hard spheres
    (icosahedral symmetry, $I_h$). The minimum distance between the outer
    spheres (os) is $r_{\rm min}^{\rm os}=\mathrm{sin}^{-1}\left(\frac{2\pi}{5}\right)
    $=1.05146222$\dots$, hence they do not touch. Bottom: The corresponding
    icosahedral graph. Numbering refers to the respective node index used in
    the Tables S1 and S2 of the Supporting Information.}
    \label{fig:GN}
\end{figure}

Newton and Gregory argued about the maximum possible number $N_k(d)$ (the {\it
maximum kissing number} or {\it Newton number}) of three-dimensional unit
spheres ($d=3$) that could be brought into contact with a central sphere
\cite{Pfender_2004}. Sch\"utte and van der Waerden provided the first proof in
1953 that max$\{N_k(3)\}=12$ \cite{Schutte_ProblemdreizehnKugeln_1952}. We call
such a cluster of 12 unit spheres kissing a central unit sphere a {\it
Gregory-Newton cluster} (GNC), shown in its most symmetric icosahedral
form in Figure \ref{fig:GN}. Exact Newton numbers for unit spheres in lattice
packings are known for dimensions $d=1$ to 9 and $d=24$, and for non-lattice
packings for $d=1-3$, 8 and 24 \cite{conway-2013book,Musin_2017}. Lower and
upper bounds for $\mathrm{max}\{N_k(d)\}$ are also available
\cite{Mittelmann_Vallentin_2010,conway-2013book}. The more general problem of
$N$ spheres of equal radius $r$ touching a given central sphere of radius 1 in
three dimensions has recently been reviewed in detail by Kusner et al.
\cite{Kusner_ConfigurationSpacesEqual_2016}. It is believed that the unit
sphere radius $r = 1$ is the maximal radius where the spheres are arbitrarily
permutable with motions remaining on the surface of a central sphere
\cite{Kusner_ConfigurationSpacesEqual_2016}. 

The Tammes, Thomson or related models employ repulsive forces between points or
spheres \cite{Wales_Ulker_2006,Wales_McKay_Altschuler_2009} and, for the
three-dimensional problem with 12 kissing spheres, lead to ideal icosahedral
symmetry (Figure \ref{fig:GN}). We may however pose the question in a slightly
different way: What happens if we let the outer kissing spheres of a GNC touch
each other to enforce rigidity? We could try to find the global and all local
minima for the 12 kissing hard spheres interacting through an attractive
instead of a repulsive potential. For example, we can place the central hard
sphere in a gravitational field of strength $F_G=Gm_im_jr_{ij}^{-2}$ and relax
all positions $r_{ij}\ge (R_i+R_j)$ between the kissing hard spheres $i$ and
$j$, in the most general case having sphere radii $R_i$ and masses $m_i$. It is
clear that such a procedure leads to a less flexible and more rigid sphere
packing. In Euclidian space, this problem is well known to
crystallization/sedimentation phenomena modelled by hard spheres in a
gravitational field \cite{Levin_2000,Pusey-2009}.

The most widely used interaction potential in chemical and physical sciences is the so-called Lennard-Jones (LJ) potential \cite{Jones_DeterminationMolecularFields_1924,Lennard-Jones_Cohesion_1931} (which includes the gravitational potential just mentioned). In reduced units the LJ$(a,b)$ potential takes the form, 
\begin{equation}
V_{a,b}^\mathrm{LJ}(r_{ij})=\frac{ar_{ij}^{-b}-br_{ij}^{-a}}{b-a} \quad ({\rm with} \ \ r_{ij},a,b \in \mathbb{R}_+ \ \ {\rm and} \ \ b>a).
\label{eqn:nmpot}
\end{equation}
It is attractive in the long range and repulsive in the short range. Such a potential maximizes the number of contacts between spheres, and for the famous LJ(6,12) case leads to one and only one minimum for the GNC \cite{Trombach_2018} -- the ideal icosahedron (shown in Figure \ref{fig:GN}) as in the case of the Tammes problem. The icosahedral motif originally proposed by Mackay \cite{Mackay-1962} plays a very important role in cluster physics and chemistry \cite{Hoare_Physicalclustermechanics_1975,Klots90,Uppenbrink-1991,vandewaal93,Wales-1996a2,wales_1996a3,Wales_Ulker_2006}.

A nice feature of the LJ potential is that for large exponents $(a,b), ~b>a$, it approaches the sticky hard-sphere (SHS) limit, originally introduced by Baxter \cite{baxter68,Gazzillo_2004},
\begin{align}
    \lim_{a,b\rightarrow \infty} V_{a,b}^\mathrm{LJ}(r_{ij}) = V_\mathrm{SHS}(r_{ij})=
    \begin{cases}
        \infty   & r_{ij} < 1\\
        -1  \quad {\rm for} & r_{ij} = 1\\
        0       & r_{ij} > 1.
    \end{cases}
\label{eqn:KS}
\end{align}
SHS models have been used intensively in many areas, such as crystallization, flocculation, colloidal suspensions, micelles, protein solutions, or in the exact enumeration of rigid SHS clusters \cite{Stell_1991,Jamnik_1996,Santos_1998,Gazzillo_2004,Hoy_MinimalEnergyPackings_2010,Arkus_Minimalenergyclusters_2009,Arkus-2010,Arkus_DerivingFiniteSphere_2011,Hoy_Structurefinitesphere_2012,Hayes_ScienceStickySpheres_2012,Holmes-Cerfon_geometricalapproachcomputing_2013,Holmes-Cerfon_EnumeratingRigidSphere_2016,Holmes-Cerfon_StickySphereClusters_2017,Kallus_Freeenergysingular_2017}. The SHS model has the advantage that an adjacency matrix $A$ can be introduced with entries $A_{ij}=1$ if spheres $i$ and $j$ touch ($r_{ij}=1$), and 0 otherwise ($r_{ij}>1$). The number of contacts between spheres then simply becomes $N_c=\sum_{i<j}^N A_{ij}$. It also opens the way for a graph-theoretical treatment of cluster structures as we shall see.

A putatively complete set of non-isomorphic rigid sphere packings (SHS
clusters) has recently been determined for cluster size $N \leq 14$ via exact
enumeration studies employing geometric rejection and rigidity rules
\cite{Hoy_Structuredynamicsmodel_2015,Holmes-Cerfon_EnumeratingRigidSphere_2016,Holmes-Cerfon_StickySphereClusters_2017}.
These include the subset of a rather large number of non-isomorphic rigid GNCs
\cite{Trombach_2018}.  In addition, the condition $\lim_{a,b\rightarrow \infty}
V_{a,b}^\mathrm{LJ}(r) = V_\mathrm{SHS}(r), b>a,$ implies that at certain $a,b$
values symmetry broken solutions away from the ideal icosahedral structure must
appear. Where exactly this happens, and when the icosahedral structure does not
survive anymore, is not known. In order to close this gap, we decided to
analyse the rigid GNCs and corresponding symmetry breaking effects in detail.
This is much in the spirit of Wales, who already pointed out that the global
characteristics of the energy landscape of a cluster can be quite sensitive to
the nature of the interatomic potential applied
\cite{Wales_MicroscopicBasisGlobal_2001}.

\section{Computational Methods}

Coordinates for GNC structures have been obtained by searching for adjacency
matrices of the results for $N=13$ from Ref.
\cite{Holmes-Cerfon_EnumeratingRigidSphere_2016} with one row or column
containing twelve "1" entries. Sub-graph isomorphism was verified using the
\textit{VF2} algorithm \cite{Cordella_SubGraphIsomorphism_2004} as implemented
in the \textit{boost graph library} \cite{_boost_2002}. Structural optimisations
with LJ potentials have been carried out using the multidimensional function
minimiser from the C++ library \textit{dlib}
\cite{King_DlibmlMachineLearning_2009} and an energy convergence criterion of
$10^{-15}$. Results from the optimisation procedure were analysed based on the
Euclidean distance matrix, which is unique for non-isomorphic structures apart from permutation, translation, rotation and inversion. For this we sorted the distances lexicographically.

\section{Results and Discussion}

\subsection{Rigid Gregory-Newton Clusters and Corresponding Graphs}

The recent results by Holmes-Cerfon
\cite{Holmes-Cerfon_EnumeratingRigidSphere_2016} contain a putatively complete
set of rigid (and few semi-rigid) SHS clusters of size $N=13$ and $N=14$ (for
details on the near completeness of the set of rigid clusters obtained see the
discussion in Ref.\cite{Holmes-Cerfon_EnumeratingRigidSphere_2016}). The rigid
GNCs can easily be identified as a subset of the set of all non-isomorphic
rigid SHS clusters, i.e. $\{ S_{\rm GN}\}\subset \{ S_{\rm SHS}\}$; these have
adjacency matrices $A$ with exactly one column and row containing twelve "1"
entries due to 12 spheres kissing the central sphere (as these cluster lie in
the region of high contact numbers with $N_c\ge 3N-6$, we expect that the set
is most likely complete).  A surprisingly large number of 737 non-isomorphic $N
= 13$ GNCs out of 98,540 rigid SHS clusters can be found \cite{Trombach_2018}.
There are four different possible contact numbers $N_c$ with $\{724,10,1,2\}$
rigid GNCs corresponding to $N_c=\{33,34,35,36\}$, therefore, non of those
clusters are hypostatic.

For further analysis and without loss of generality we delete the central
sphere and analyse the remaining non-isomorphic shell of spheres (note that
rigidity requires the presence of the central sphere), also called contact
graphs according to Sch{\"u}tte, van der Waerden and Habicht
\cite{Schutte_1951}. This has the advantage that these shells are related to
planar connected graphs. In the following we call the corresponding connected,
planar graph of such a shell of spheres with the central sphere missing a {\it
GN graph}. The question arises if all 737 non-isomorphic GN graphs are
subgraphs to the icosahedral graph, as shown in Figure~\ref{fig:GN}. This would
make sense as it is impossible to increase the degree of any vertex beyond 5 in
the GN graph. Note that the icosahedral cluster is completely unjammed and its
space of (infinitesimal) deformations has dimension 24 (for details see
Ref.\cite{Kusner_ConfigurationSpacesEqual_2016}).

Employing the \textit{VF2} algorithm \cite{Cordella_SubGraphIsomorphism_2004}
as implemented in the \textit{boost graph library} \cite{_boost_2002} we find
all 737 non-isomorphic GN graphs $G_{\rm GN}(V,E')$ (vertex count $|V|=12$,
edge count $|E'|<30$) to be (edge-induced) subgraphs of the icosahedral graph
$G_{\rm ico}(V,E)$ ($|V|=12$, $|E|=30$), which implies that their vertices can
all be mapped to vertices of the icosahedral graph with certain edges deleted
such that the subgraph remains connected ($V_{\rm GN} = V_{\rm ico}$ and
$E_{\rm GN}\subset E_{\rm ico}$). An extensive list of all subgraphs is
included in the Supporting Information (Tables S1 and S2). Note, not all GN
graphs are 3-connected and therefore are not strictly polyhedral according to
Steinitz's theorem \cite{Steinitz1922}. These are the graphs which have
vertices of degree 2, i.e. $|V_2|>0$, and there are 304 of them, Table~S1. As
the many non-isomorphic graphs listed in the SI are obtained from a certain
combination of edge deletions under the constraint of maintaining rigidity, it
is not surprising at all that the number of non-isomorphic GN graphs is so
large.

\begin{figure}[htb]
    \centering
    \subfloat[hcp, $|E|=24, \omega =1$.\label{subfig:hcpgraph}]{\includegraphics[width=0.5\columnwidth]{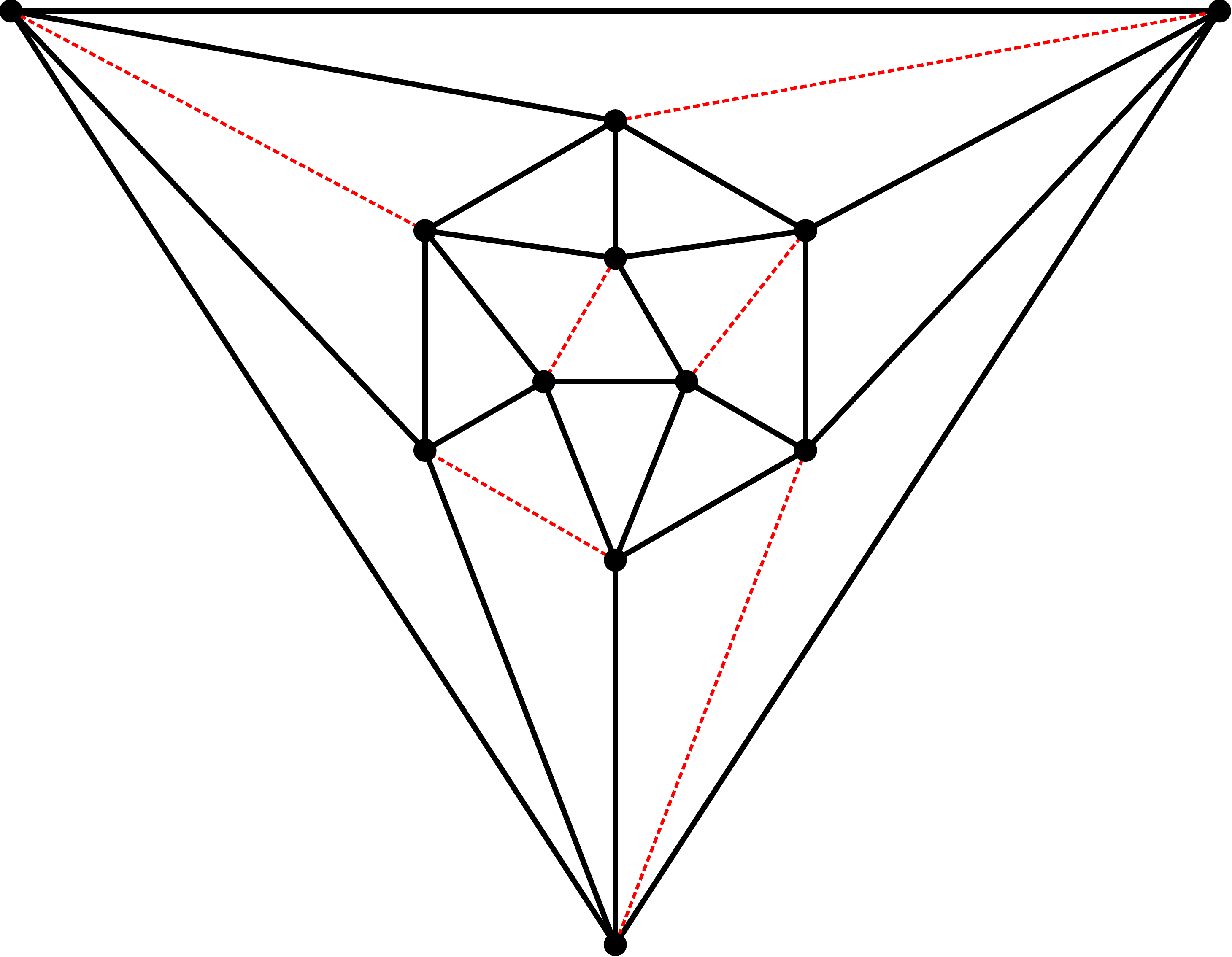}\hspace{0.03\textwidth}\includegraphics[width=.42\columnwidth]{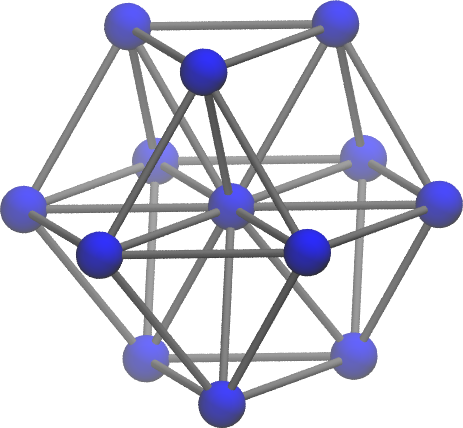}}\\
    \subfloat[fcc, $|E|=24, \omega =2$.\label{subfig:fccgraph}]{\includegraphics[width=0.5\columnwidth]{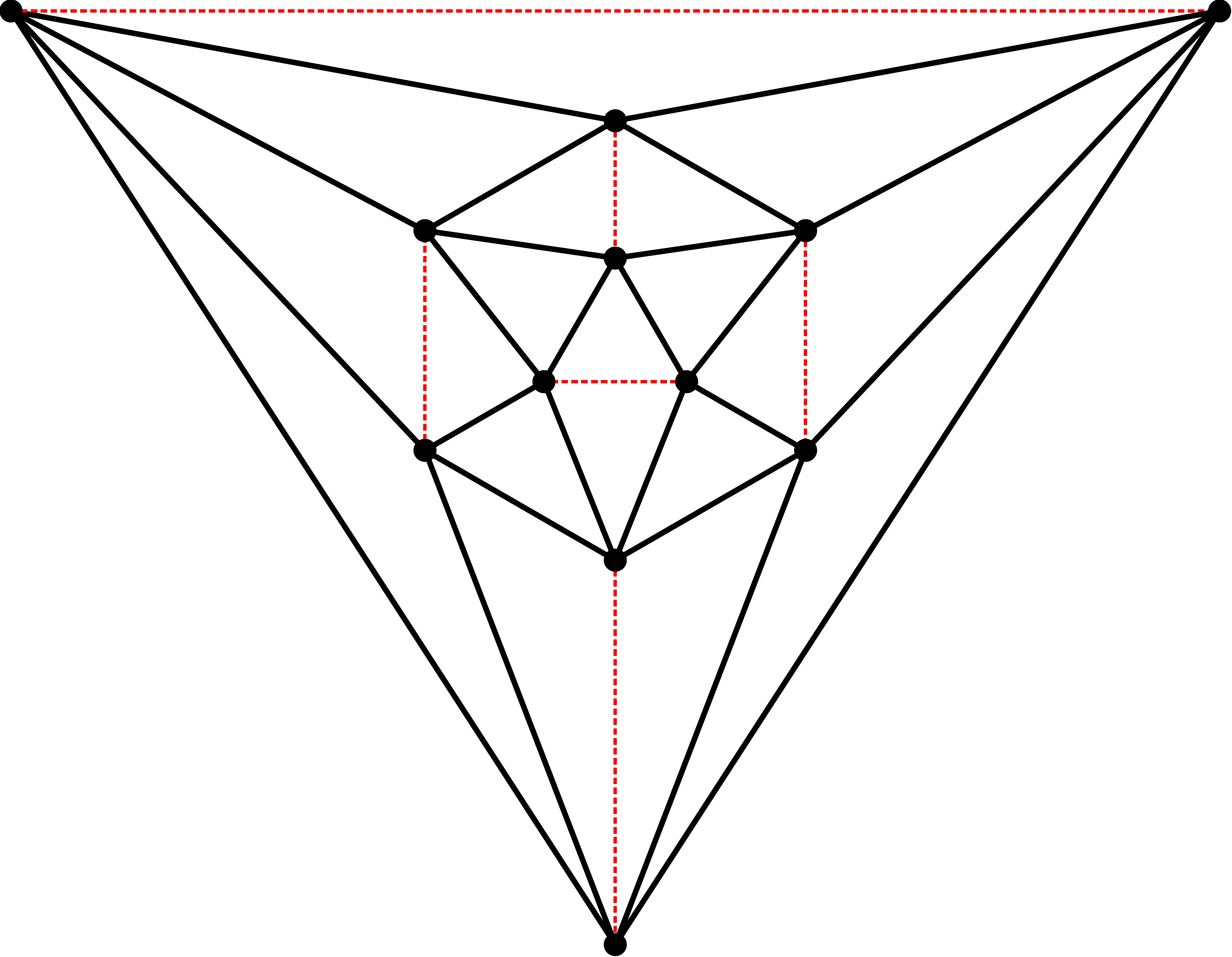}\hspace{0.03\textwidth}\includegraphics[width=.42\columnwidth]{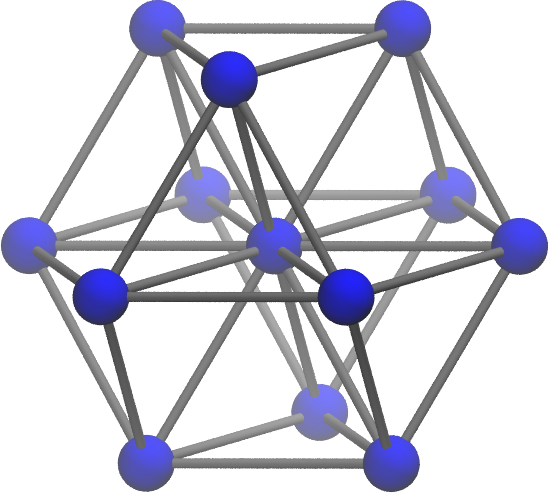}}
    \caption{GN hcp (triangular orthobicupola) and fcc (cuboctahedron) graphs
    (central sphere removed) as subgraphs of the icosahedral graph and
    corresponding rigid GNCs.  Red lines indicate the edges that were removed
    to create the GN graph. The ordinal numbers $\omega$ refer to Table~S2 in
    the SI.}
    \label{fig:GNshellgraphs}
\end{figure}

The results show, that at least six and up to a maximum of nine edges have to
be removed from the icosahedral graph to create a GN graph. Removing six edges
from the icosahedral graph results in 24 edges, or $N_c=36$ if we include the
central sphere. For $N=13$ this is exactly equal to $3N-3$ which is the maximum
contact number observed for this cluster size
\cite{Hoy_Structuredynamicsmodel_2015,Holmes-Cerfon_EnumeratingRigidSphere_2016}.
Consequently, removing nine edges gives $N_c=33=3N-6$, meaning that rigid GNCs
cannot be hypostatic (i.e. $N_c < 3N-6$). Interestingly, there are only two
graphs with maximum edge count of $|E|=24$, which are exactly the fragments of
the face-centered cubic (fcc, ABCABC... layers) and hexagonal closed packed
(hcp, ABAB... layers) bulk structures, respectively. These are the result from
removing 6 edges in such a way, that exactly one edge is removed from every
vertex in the icosahedral graph (thus the degree of every vertex is 4), see
Figure~\ref{fig:GNshellgraphs}.
Removing edges in this way implies that the resulting two graphs consist of
triangles and rectangles only. The difference between the fcc and hcp clusters
is in the way their square faces are connected; in the fcc case the square
faces only connect via edges (cuboctahedron), while in hcp case the square
faces come in pairs sharing one edge (triangular orthobicupola or Johnson solid
$J_{27}$) \cite{Kusner_ConfigurationSpacesEqual_2016}.

\begin{figure}[htb]
    \centering
    \includegraphics[width=\columnwidth]{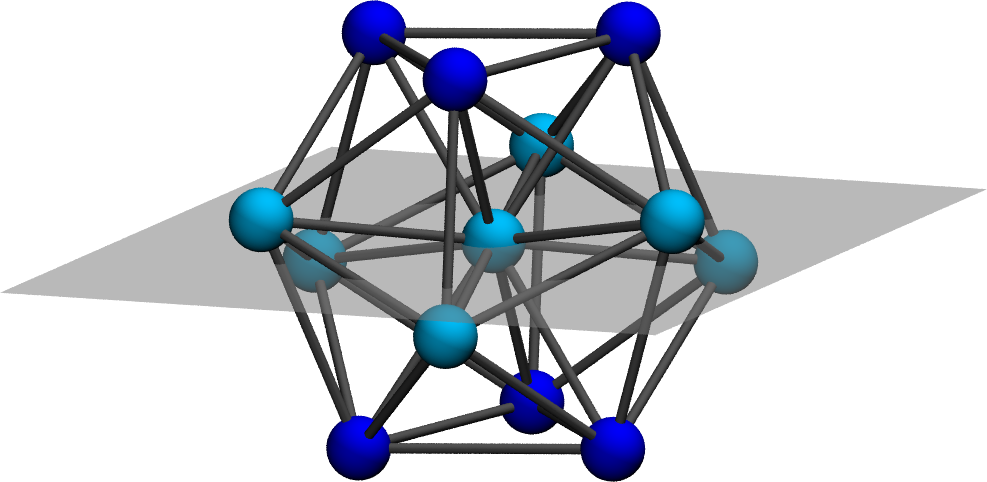}
    \caption{Illustration of one zig-zag path (light blue spheres) that needs
    to be deformed such that it aligns with the triangular plane (shown in
    grey) of the fcc crystal.}
    \label{fig:ico-fcc-trans}
\end{figure}

The construction of hcp and fcc structures by a continuous deformation of an
icosahedron has been described in detail by Kusner et al.
\cite{Kusner_ConfigurationSpacesEqual_2016} and goes back to Conway and Sloane
in 1988 \cite{conway-2013book}. We note that hcp and fcc can both be obtained
from a rearrangement of the spheres in an icosahedron by forming a (zig-zag)
cycle (closed path) through six vertices, and arranging those spheres on the
path such that they are in-plane with the central sphere, which becomes part of
the hexagonal plane as in the bulk fcc and hcp packing
(Figure~\ref{fig:ico-fcc-trans}).
Additionally, the plane has to be rotated by $\pi/6$ to create the fcc
structure. The hcp structure can be constructed by also rotating either the top
or the bottom plane by the same amount in either direction parallel to the
hexagonal plane. Kusner noted that a smooth deformation from the icosahedral
configuration to hcp requires 9 moving spheres
\cite{Kusner_ConfigurationSpacesEqual_2016}. This interesting transition path
may be the key for the icosahedral to closed-packed rearrangements in larger
clusters, which has previously been described in terms of catastrophe theory as
a cusp catastrophe \cite{Wales_MicroscopicBasisGlobal_2001}. 

\begin{figure}[htb]
    \centering
    \subfloat[icosahedral motif, $|E|=22, \omega =4$.\label{subfig:ico4graph}]{\includegraphics[width=0.5\columnwidth]{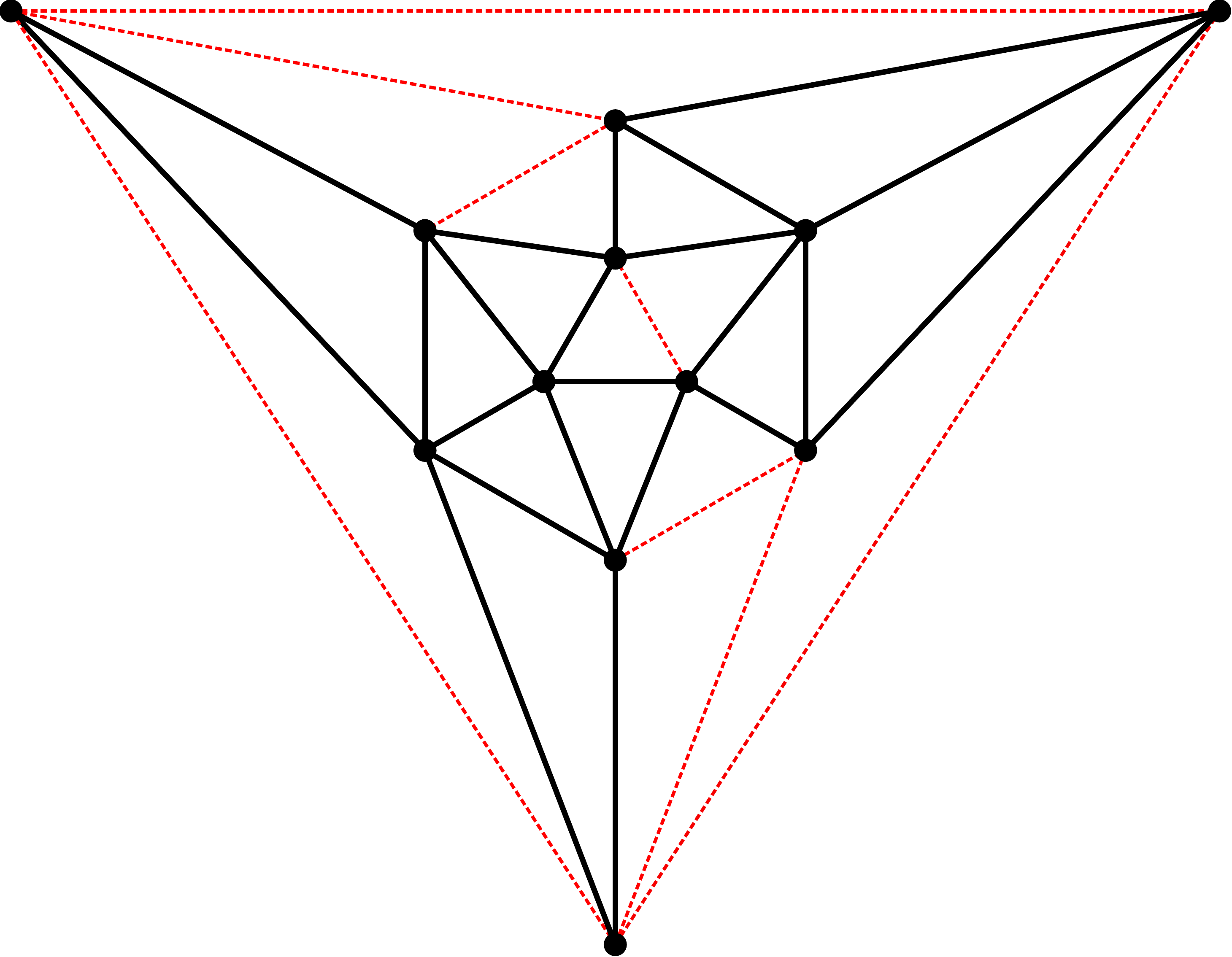}\hspace{0.03\textwidth}\includegraphics[width=.32\columnwidth]{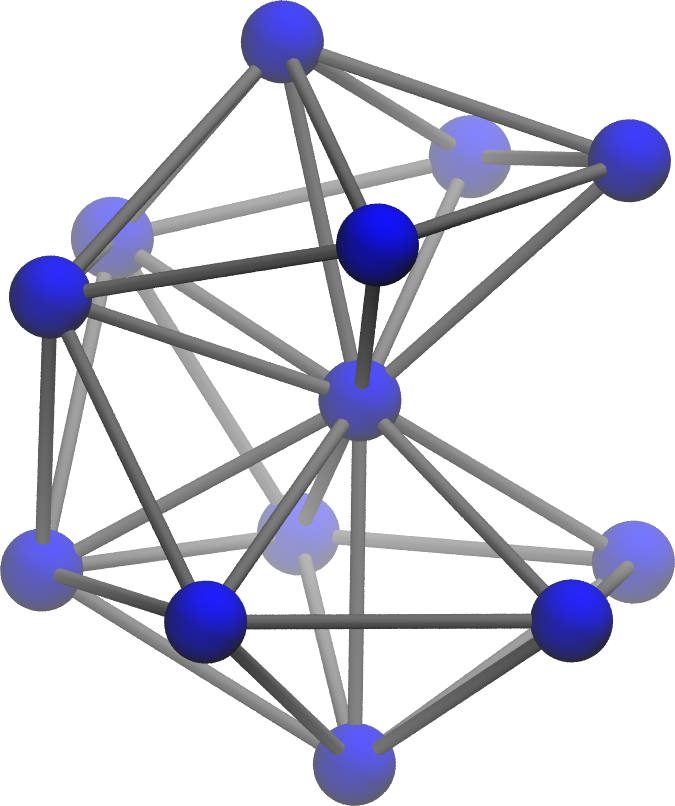}}\\
    \subfloat[icosahedral motif, $|E|=22, \omega =7$.\label{subfig:ico7graph}]{\includegraphics[width=0.5\columnwidth]{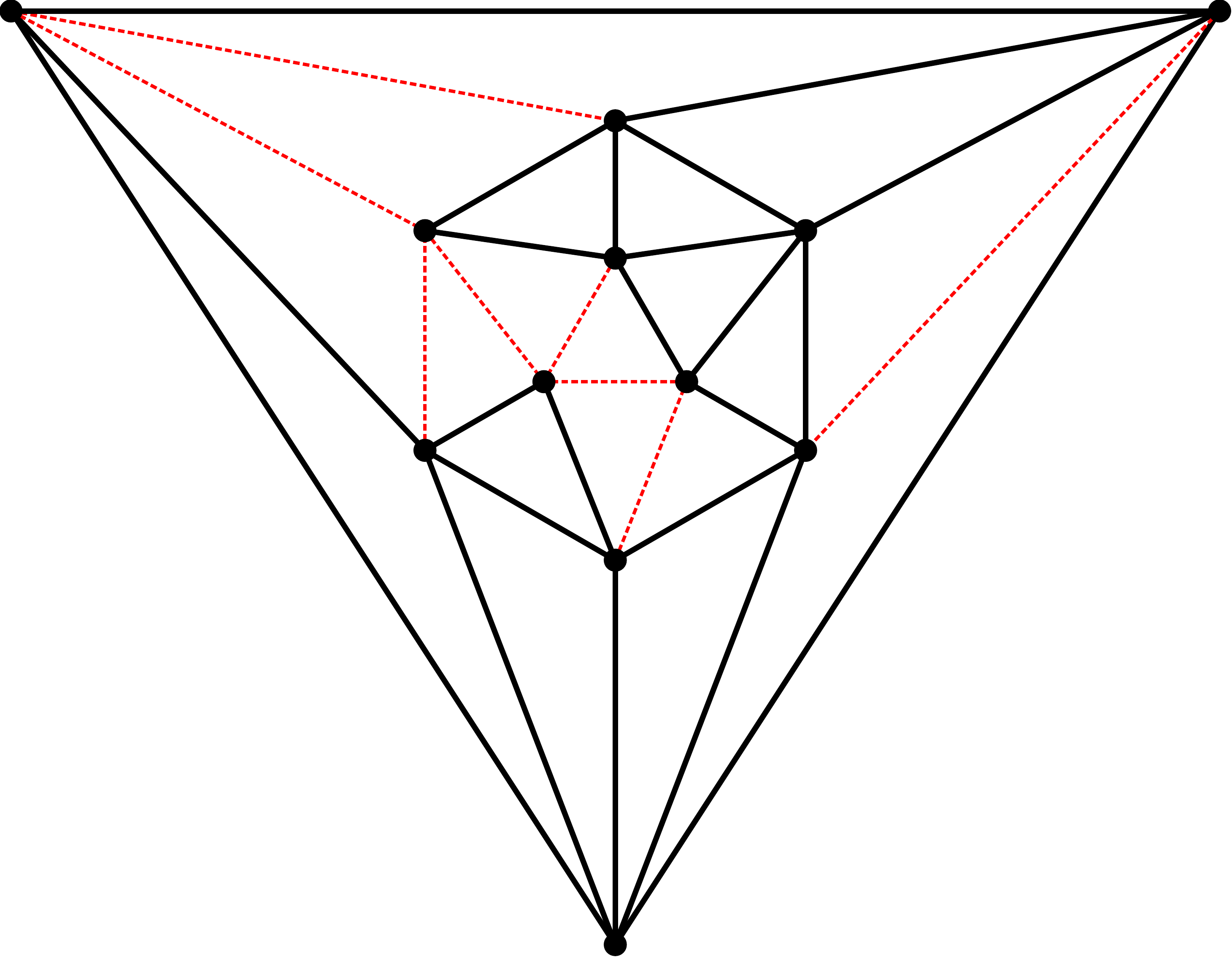}\hspace{0.03\textwidth}\includegraphics[width=.32\columnwidth]{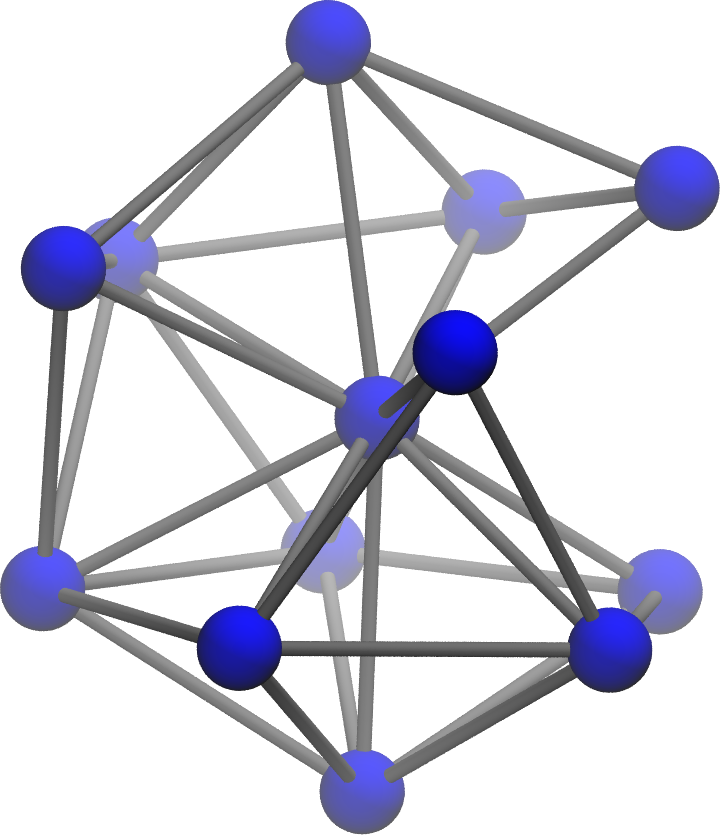}}
    \caption{Representative GN graphs (central sphere removed) with $|F_3|=10$
    as subgraphs of the icosahedral graph and corresponding rigid GNCs. The
    icosahedral motif in the 3D embedding is clearly visible.  Red lines
    indicate the edges that were removed to create the GN graph. The ordinal
    numbers $\omega$ refer to Table~S2 in the Supporting Information.}
    \label{fig:GNicographs}
\end{figure}

Even though the rearrangement from the icosahedral to either the fcc or hcp
cluster structure can easily be realized for the GNC, there should be clusters
where the icosahedral motif is still clearly visible, i.e. only small
rearrangements of the spheres are necessary to break icosahedral symmetry and form
a rigid cluster. These are, for example, the ones with maximum count of
triangles, i.e. according to Table~S1 (SI) the GN graphs with $|F_3|=10$ with
edge counts of $|E|=$ 22 or 21. Two of these are shown with their corresponding
graphs in Figure \ref{fig:GNicographs}.

\begin{figure}[htb]
    \centering
    \subfloat[Distorted elongated pentagonal bipyramid (Johnson solid $J_{16}$), $|E|=23, \omega =3$.\label{subfig:johnsongraph}]{\includegraphics[width=0.5\columnwidth]{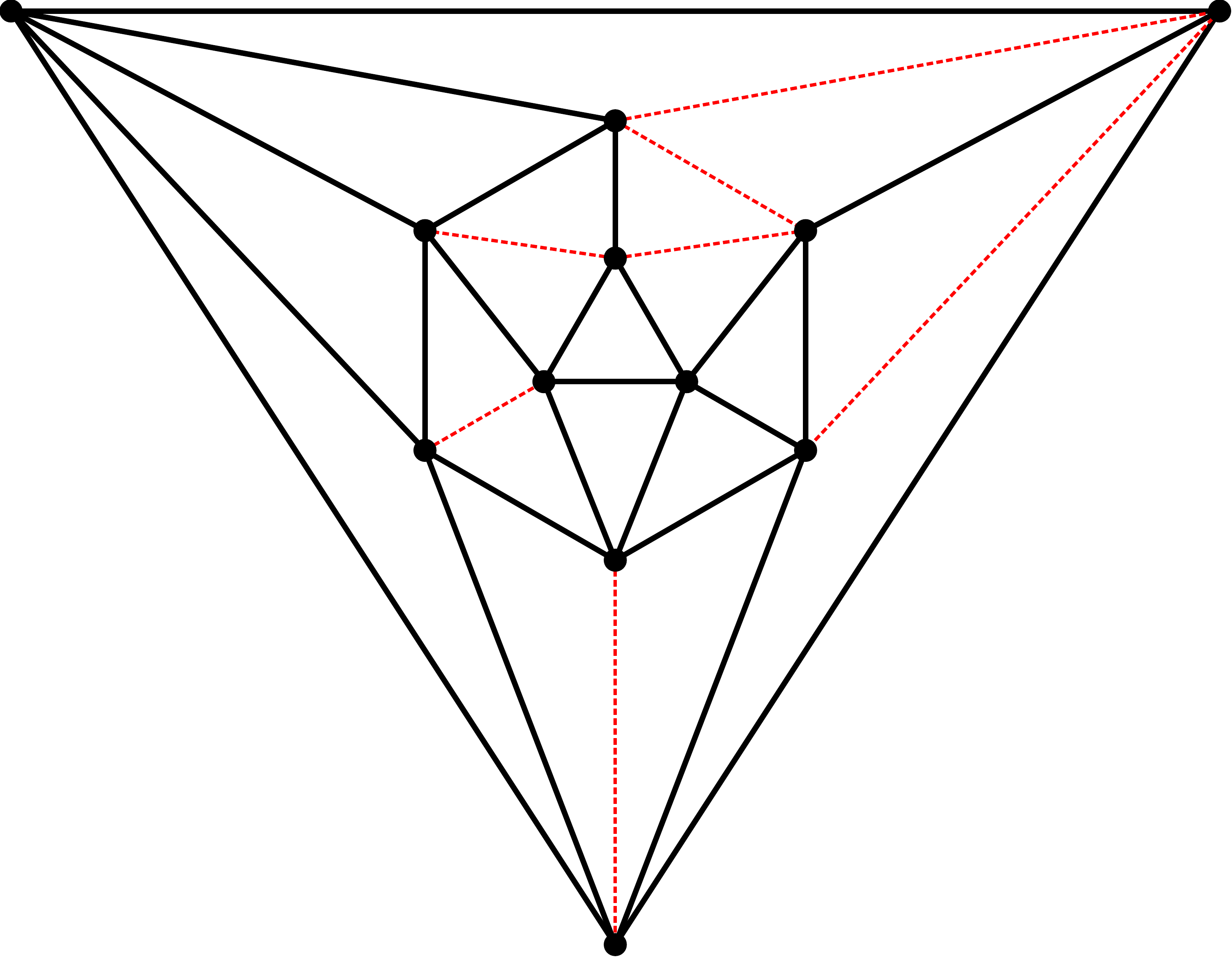}\hspace{0.03\textwidth}\includegraphics[width=.32\columnwidth]{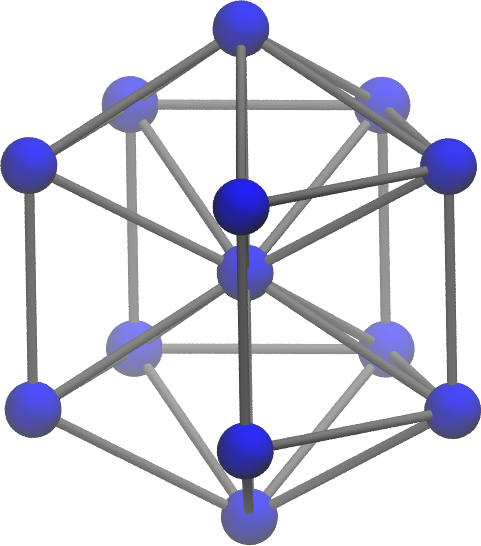}}
    \caption{GN graph (central sphere removed) as subgraphs of the icosahedral graph and corresponding 
    GN Johnson-like solid (with edges removed). Red lines indicate the edges that were removed 
    from the icosahedral graph to create the GN graph.
    The ordinal number $\omega$ refers to Table~S2 in the Supporting Information.}
    \label{fig:GNJohnsongraph}
\end{figure}

Figure \ref{fig:GNJohnsongraph} shows the graph with the next highest edge
count after the fcc and hcp packings. The motif of a distorted elongated
pentagonal bipyramid (Johnson solid $J_{16}$) is clearly visible. Note that the
Johnson solid can be obtained by deleting five edges in the icosahedral graph
and rotating the two
opposite pentagonal pyramids by $2\pi /5$. One of the resulting square faces
has to be stretched to obey the SHS conditions, which is achieved by removing
two additional edges. In the graph this implies that a hexagonal face is
formed. Note that this GNC is also the cluster with the largest distance
$r_\mathrm{max}^\mathrm{RE}= 1.47823719$ that corresponds to a removed edge
(RE) in the GN graph. Capping this cluster with one more sphere over the
distorted square face with $r_\mathrm{max}^\mathrm{RE}$ leads to the structure
with the shortest distance to the central sphere a sphere in the second 
coordination shell can have ($r^\mathrm{COS}=1.347150628$) out of all 14,529 GN
clusters with $N=14$ \cite{Trombach_2018}.

If more edges are removed from the icosahedral graph we see the appearance of
larger $n$-gonal faces with the largest face being a 12-gon.

\subsection{Symmetry-Broken Lennard-Jones Gregory-Newton Clusters}

\begin{figure}[htb]\centering
    \includegraphics[width=\columnwidth]{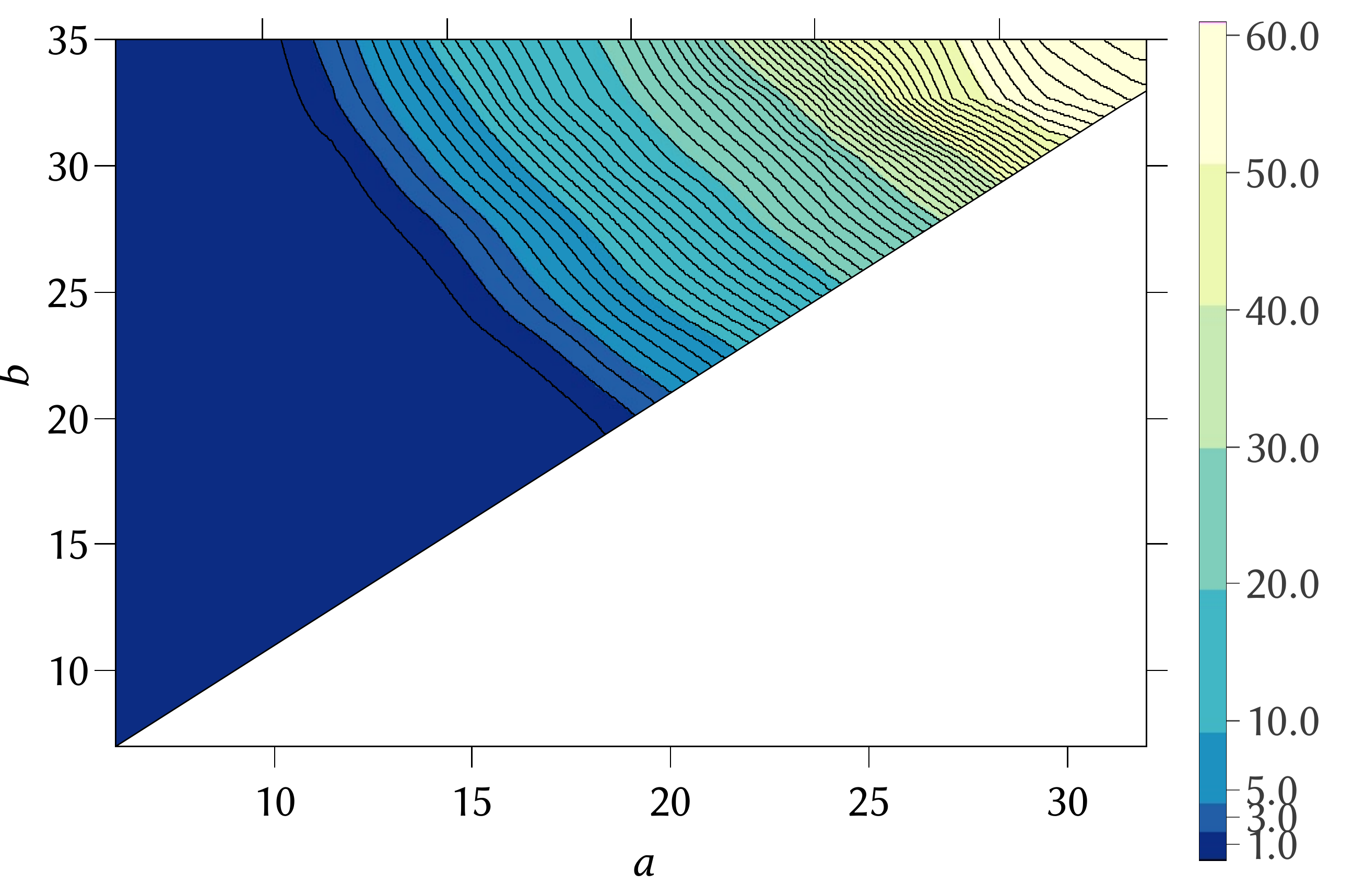}
    \caption{Number of unique structures resulting from an optimisation with a
    LJ$(a,b)$ potential. The lowest contour line shows the point where more
    than one structure results from the optimisation and the distance between
    contour lines is 1.}
    \label{fig:ico-2d}
\end{figure}

All 737 non-isomorphic rigid GNCs optimise to the ideal icosahedral symmetry if
a LJ(6,12) potential is applied \cite{Trombach_2018} (however, for larger sized
icosahedral structures many more minima appear, see Refs.
\cite{Doye_1995,Wales_1996a1,Doye_1996a,Doye_1997a}). As mentioned in the
introduction, for equally sized hard spheres a cluster with icosahedral
symmetry leaves gaps between the spheres on the outer shell, i.e. they do not
touch, and is therefore not considered rigid under SHS conditions. Hence, at
certain $(a,b)$ combinations a phase transition must occur in the LJ$(a,b)$
energy landscape where local minima appear, which do not have icosahedral
symmetry anymore. In order to determine those $(a,b)$ combinations, we
optimised all 3D cluster geometries with varying exponents $(6\leq a \leq 34$
and $7\leq b\leq 35)$ with $(b>a)$ and analysed the number of resulting minimum
structures. The results are shown schematically in Figure~\ref{fig:ico-2d}.

\begin{figure}[htb]\centering
    \includegraphics[width=\columnwidth]{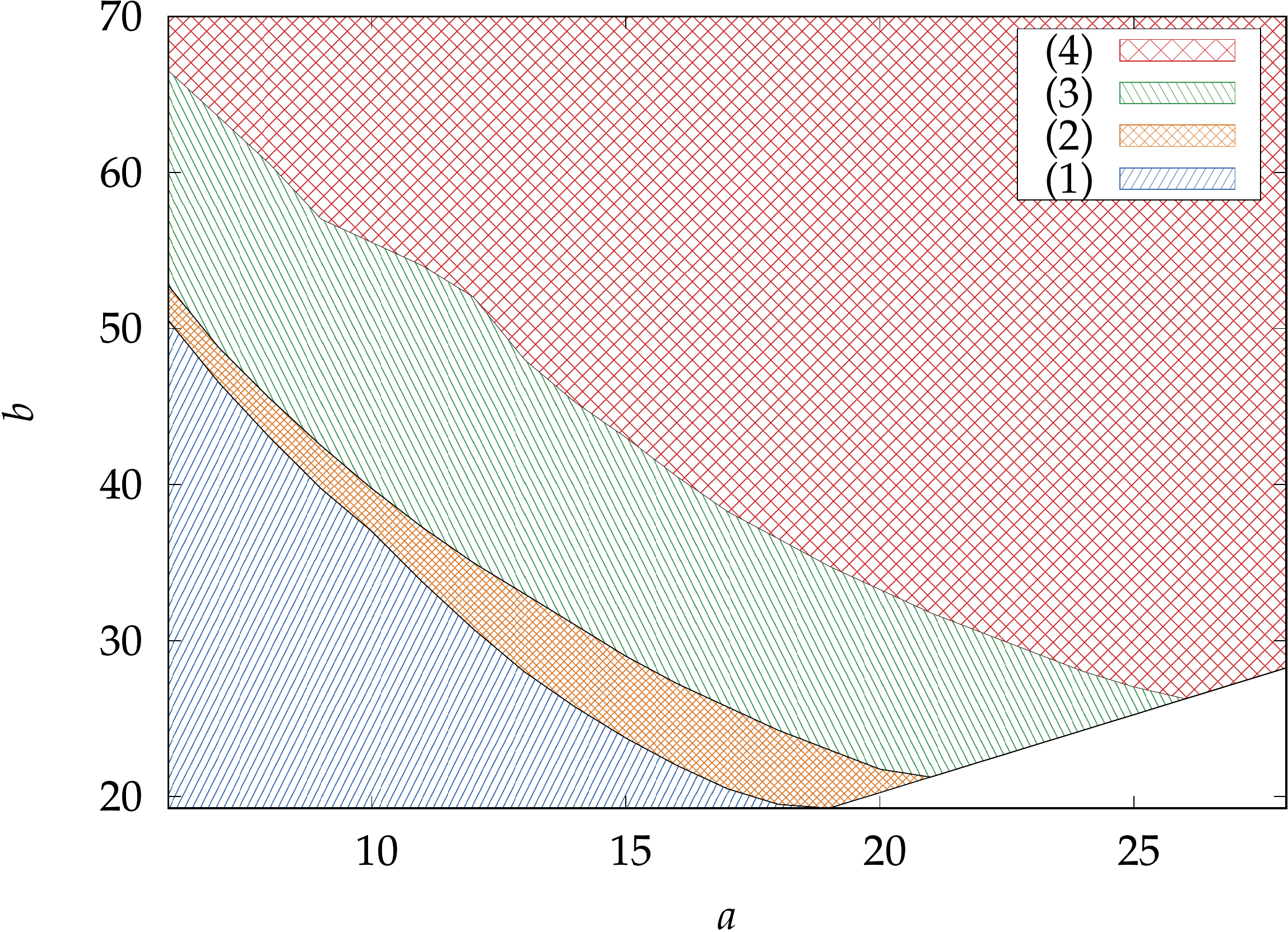}
    \caption{Different types of energy landscapes arising from combinations of
    the LJ $(a,b)$ exponents. (1) One single (icosahedral) minimum, (2) more
    than one minimum with the icosahedron as the global minimum, (3) more than
    one minimum with the icosahedron becoming a local (and not global) minimum,
    (4) the icosahedral motif disappears completely. The unshaded small area
    in the bottom right corner corresponds to $a>b$, which is excluded. The
    resolution for $a$ is 1.0 and for $b$ 0.25.}
    \label{fig:no-ico}
\end{figure}

Figure~\ref{fig:no-ico} contains additional information showing three major
phase transitions in the topology of the energy landscape going from low to
high $(a,b)$ exponents. In the blue shaded area (1), the Mackay icosahedron is
the sole minimum in the potential energy landscape. The first transition occurs
when this symmetry can be broken, and other local minima are supported by the
LJ$(a,b)$ potential besides the icosahedron. This is indicated in
Figure~\ref{fig:no-ico} by the smallest, orange region (2), which still
contains the perfect icosahedron as the global minimum. At slightly higher
exponents, other structures become energetically more favourable and replace
the icosahedron as the global minimum, region (3). However, the icosahedron
remains as a local minimum in the potential energy surface. The last transition
occurs when the LJ potential becomes SHS-like, and the icosahedral cluster
completely disappears from the potential energy surface, region (4). The three
transition lines are generally smooth.

\begin{figure}[htb]\centering
    \includegraphics[width=\columnwidth]{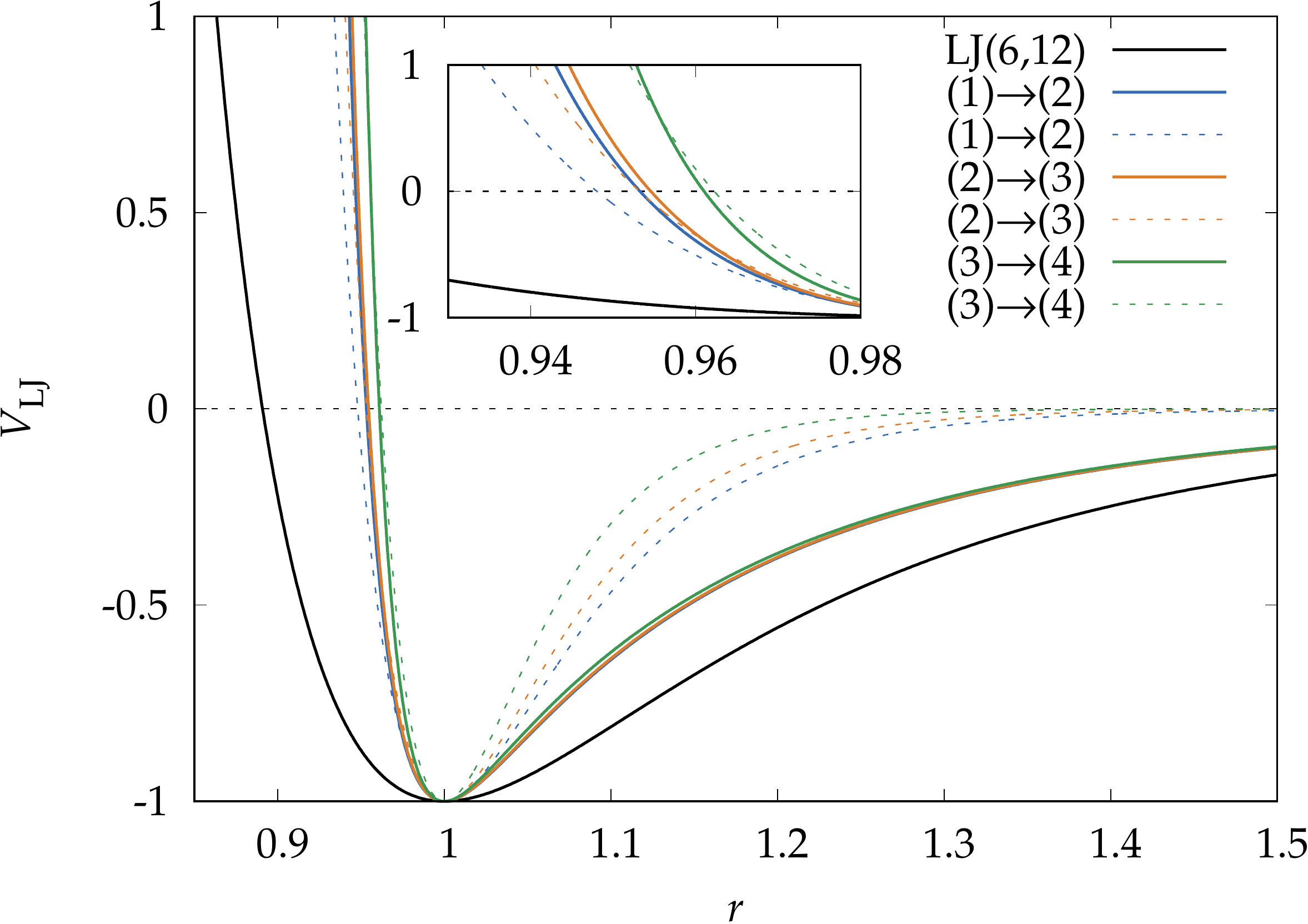}
    \caption{Comparison of different shapes of LJ potentials at the phase
    transition lines shown in fig.~\ref{fig:no-ico} with the traditional
    LJ(6,12) potential (black solid line). Dashed lines refer to potentials
    with low $a$ values (left side of fig.~\ref{fig:no-ico}), while solid lines
    refer to potentials with high $a$ values (right side of
    fig.~\ref{fig:no-ico}).}
    \label{fig:compareLJ}
\end{figure}

Figure~\ref{fig:compareLJ} shows representative LJ potentials for combinations
of the $(a,b)$ exponents (with low and high $a$ values) on the phase transition
lines drawn in figure~\ref{fig:no-ico}. At these phase transition lines, the
corresponding LJ potentials show narrow and steep repulsive potentials compared
to the LJ$(6,12)$ potential, which all look very similar in the short range ($r<1$).
However, they differ substantially in the long range ($r>1$).

\begin{figure}[htb]\centering
    \includegraphics[width=\columnwidth]{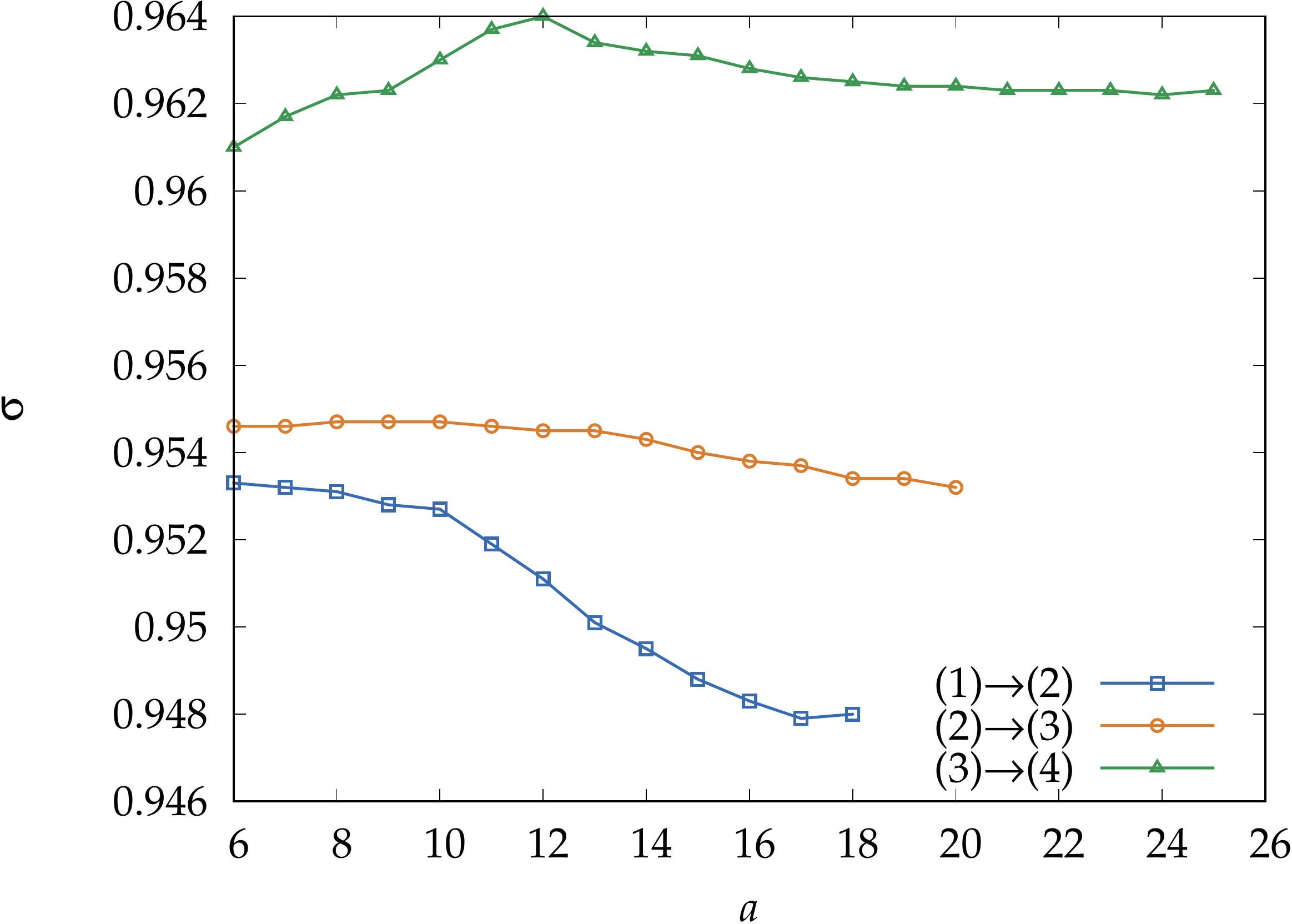}
    \caption{Hard-sphere radii $\sigma$ in reduced units for the LJ$(a,b)$
    potentials on the transition lines shown in fig.~\ref{fig:no-ico}.}
    \label{fig:hardsphere}
\end{figure}

The $(a,b)$ parameters can be related to the so-called LJ hard-sphere radius $\sigma$ (given by the
intersection with the abscissa) through equation~\eqref{eqn:nmpot}, 
\begin{equation}
    \sigma=\left(\frac{b}{a}\right)^{\frac{1}{a-b}}.
\end{equation}
and we only have to consider the $(a,\sigma)$ combinations shown in  
Figure~\ref{fig:hardsphere} along the phase transition lines.

The variation in $\sigma$ along the phase transitions lines for (2)$\rightarrow$(3)
and (3)$\rightarrow$(4) are rather small. However, all three
transitions clearly show different ranges for $\sigma$ and thus can be
characterized by the LJ hard-sphere radius. These are also
much larger compared to the LJ(6,12) hard-sphere radius of $\sigma=0.891$,
and close to the ideal hard sphere radius of 1 within the SHS model. 
This demonstrates that the shape of the LJ potential in the
repulsive region has a significant influence on the position of the transition lines, 
and therefore on the topology of the energy landscape. In contrast, these 
transitions seem to be far less affected by the shape of the potential in the attractive region. 
Only for the transition (1)$\rightarrow$(2) we see a larger variation in $\sigma$.

Of course, symmetry breaking (such as Jahn-Teller distortions) in icosahedral
clusters is well known in cluster chemistry and physics \cite{Heer-1993},
which, however, requires the introduction of many-body forces beyond the usual
two-body interaction, or a correct quantum theoretical treatment containing all
many-body forces. 

\section{Conclusions}

We have analysed rigid GNCs by graph theoretical means. All 737
non-isomorphic GN graphs are subgraphs of the icosahedral graph obtained by
deleting a minimum of 6 and a maximum of 9 edges. There are only two structures
with maximum edge count of 24 corresponding to the sphere packing of the fcc
and hcp structures, which can be obtained from the icosahedral structure by a
smooth rearrangement moving the six spheres along a closed zig-zag path into
the (hexagonal) plane. The common LJ(6,12) potential has only one minimum
structure corresponding to the ideal icosahedron where the 12 outer spheres do
not touch each other. Symmetry breaking requires a very repulsive short-range
LJ potential. We also determined the $(a,b)$-line in the LJ$(a,b)$ potential
where the icosahedron completely disappears. While our results depend on the
functional form chosen (the Lennard-Jones potential), we expect similar results for
other well known potentials such as the Morse potential.

The sphere kissing problem in
higher dimensions is a well known problem \cite{conway-2013book} (in two
dimensions there is only 1 non-isomorphic GNC). How many non-isomorphic
rigid GNCs there are in greater than three dimensions is currently
unknown. Moreover, the rigid kissing sphere problem can be extended to other
(convex or not) topologies instead of a central sphere, e.g. kissing spheres on an ellipsoid.
There are many open questions in this field.\\

\section{Acknowledgements}
We acknowledge financial support, including covering the costs to publish in open access, by the Marsden Fund of the Royal Society of New Zealand and the Centre for Advanced Study at the Norwegian Academy of Science and Letters (Molecules in Extreme Environments Research Program). We thank Profs. David J. Wales (Cambridge) and Robert S. Hoy (Tampa) for valuable comments.

\bibliography{GN}

\begin{thebibliography}{50}%
\makeatletter
\providecommand \@ifxundefined [1]{%
 \@ifx{#1\undefined}
}%
\providecommand \@ifnum [1]{%
 \ifnum #1\expandafter \@firstoftwo
 \else \expandafter \@secondoftwo
 \fi
}%
\providecommand \@ifx [1]{%
 \ifx #1\expandafter \@firstoftwo
 \else \expandafter \@secondoftwo
 \fi
}%
\providecommand \natexlab [1]{#1}%
\providecommand \enquote  [1]{``#1''}%
\providecommand \bibnamefont  [1]{#1}%
\providecommand \bibfnamefont [1]{#1}%
\providecommand \citenamefont [1]{#1}%
\providecommand \href@noop [0]{\@secondoftwo}%
\providecommand \href [0]{\begingroup \@sanitize@url \@href}%
\providecommand \@href[1]{\@@startlink{#1}\@@href}%
\providecommand \@@href[1]{\endgroup#1\@@endlink}%
\providecommand \@sanitize@url [0]{\catcode `\\12\catcode `\$12\catcode
  `\&12\catcode `\#12\catcode `\^12\catcode `\_12\catcode `\%12\relax}%
\providecommand \@@startlink[1]{}%
\providecommand \@@endlink[0]{}%
\providecommand \url  [0]{\begingroup\@sanitize@url \@url }%
\providecommand \@url [1]{\endgroup\@href {#1}{\urlprefix }}%
\providecommand \urlprefix  [0]{URL }%
\providecommand \Eprint [0]{\href }%
\providecommand \doibase [0]{http://dx.doi.org/}%
\providecommand \selectlanguage [0]{\@gobble}%
\providecommand \bibinfo  [0]{\@secondoftwo}%
\providecommand \bibfield  [0]{\@secondoftwo}%
\providecommand \translation [1]{[#1]}%
\providecommand \BibitemOpen [0]{}%
\providecommand \bibitemStop [0]{}%
\providecommand \bibitemNoStop [0]{.\EOS\space}%
\providecommand \EOS [0]{\spacefactor3000\relax}%
\providecommand \BibitemShut  [1]{\csname bibitem#1\endcsname}%
\let\auto@bib@innerbib\@empty
\bibitem [{\citenamefont {Tammes}(1930)}]{tammes_1930}%
  \BibitemOpen
  \bibfield  {author} {\bibinfo {author} {\bibfnamefont {P.~M.~L.}\
  \bibnamefont {Tammes}},\ }\href@noop {} {\bibfield  {journal} {\bibinfo
  {journal} {Recueil des travaux botaniques n{\'e}erlandais}\ }\textbf
  {\bibinfo {volume} {27}},\ \bibinfo {pages} {1} (\bibinfo {year}
  {1930})}\BibitemShut {NoStop}%
\bibitem [{\citenamefont {Robinson}(1961)}]{Robinson_1961}%
  \BibitemOpen
  \bibfield  {author} {\bibinfo {author} {\bibfnamefont {R.~M.}\ \bibnamefont
  {Robinson}},\ }\href {\doibase 10.1007/BF01396539} {\bibfield  {journal}
  {\bibinfo  {journal} {Math. Ann.}\ }\textbf {\bibinfo {volume} {144}},\
  \bibinfo {pages} {17} (\bibinfo {year} {1961})}\BibitemShut {NoStop}%
\bibitem [{\citenamefont {Musin}\ and\ \citenamefont
  {Tarasov}(2015)}]{Musin_2015}%
  \BibitemOpen
  \bibfield  {author} {\bibinfo {author} {\bibfnamefont {O.~R.}\ \bibnamefont
  {Musin}}\ and\ \bibinfo {author} {\bibfnamefont {A.~S.}\ \bibnamefont
  {Tarasov}},\ }\href {\doibase 10.1080/10586458.2015.1022842} {\bibfield
  {journal} {\bibinfo  {journal} {Exp. Math.}\ }\textbf {\bibinfo {volume}
  {24}},\ \bibinfo {pages} {460} (\bibinfo {year} {2015})},\ \Eprint
  {http://arxiv.org/abs/https://doi.org/10.1080/10586458.2015.1022842}
  {https://doi.org/10.1080/10586458.2015.1022842} \BibitemShut {NoStop}%
\bibitem [{\citenamefont {Pfender}\ and\ \citenamefont
  {Ziegler}(2004)}]{Pfender_2004}%
  \BibitemOpen
  \bibfield  {author} {\bibinfo {author} {\bibfnamefont {F.}~\bibnamefont
  {Pfender}}\ and\ \bibinfo {author} {\bibfnamefont {G.~M.}\ \bibnamefont
  {Ziegler}},\ }\href@noop {} {\bibfield  {journal} {\bibinfo  {journal}
  {Notices - Am. Math. Soc.}\ }\textbf {\bibinfo {volume} {51}},\ \bibinfo
  {pages} {873} (\bibinfo {year} {2004})}\BibitemShut {NoStop}%
\bibitem [{\citenamefont {Sch\"utte}\ and\ \citenamefont {van~der
  Waerden}(1952)}]{Schutte_ProblemdreizehnKugeln_1952}%
  \BibitemOpen
  \bibfield  {author} {\bibinfo {author} {\bibfnamefont {K.}~\bibnamefont
  {Sch\"utte}}\ and\ \bibinfo {author} {\bibfnamefont {B.~L.}\ \bibnamefont
  {van~der Waerden}},\ }\href {\doibase 10/c2tnph} {\bibfield  {journal}
  {\bibinfo  {journal} {Math. Ann.}\ }\textbf {\bibinfo {volume} {125}},\
  \bibinfo {pages} {325} (\bibinfo {year} {1952})}\BibitemShut {NoStop}%
\bibitem [{\citenamefont {Conway}\ and\ \citenamefont
  {Sloane}(2013)}]{conway-2013book}%
  \BibitemOpen
  \bibfield  {author} {\bibinfo {author} {\bibfnamefont {J.~H.}\ \bibnamefont
  {Conway}}\ and\ \bibinfo {author} {\bibfnamefont {N.~J.~A.}\ \bibnamefont
  {Sloane}},\ }\href@noop {} {\emph {\bibinfo {title} {Sphere Packings,
  {{Lattices}} and {{Groups}}}}},\ \bibinfo {edition} {3rd}\ ed.,\ \bibinfo
  {series} {Grundlehren der mathematischen Wissenschaften}\ No.\ \bibinfo
  {number} {290}\ (\bibinfo  {publisher} {{Springer Science \& Business
  Media}},\ \bibinfo {year} {2013})\BibitemShut {NoStop}%
\bibitem [{\citenamefont {Musin}(2017)}]{Musin_2017}%
  \BibitemOpen
  \bibfield  {author} {\bibinfo {author} {\bibfnamefont {O.~R.}\ \bibnamefont
  {Musin}},\ }\href@noop {} {\bibfield  {journal} {\bibinfo  {journal} {arXiv
  preprint arXiv:1712.04099}\ } (\bibinfo {year} {2017})}\BibitemShut {NoStop}%
\bibitem [{\citenamefont {Mittelmann}\ and\ \citenamefont
  {Vallentin}(2010)}]{Mittelmann_Vallentin_2010}%
  \BibitemOpen
  \bibfield  {author} {\bibinfo {author} {\bibfnamefont {H.}~\bibnamefont
  {Mittelmann}}\ and\ \bibinfo {author} {\bibfnamefont {F.}~\bibnamefont
  {Vallentin}},\ }\href@noop {} {\bibfield  {journal} {\bibinfo  {journal}
  {Exp. Math.}\ }\textbf {\bibinfo {volume} {19}},\ \bibinfo {pages} {175}
  (\bibinfo {year} {2010})}\BibitemShut {NoStop}%
\bibitem [{\citenamefont {Kusner}\ \emph {et~al.}(2016)\citenamefont {Kusner},
  \citenamefont {Kusner}, \citenamefont {Lagarias},\ and\ \citenamefont
  {Shlosman}}]{Kusner_ConfigurationSpacesEqual_2016}%
  \BibitemOpen
  \bibfield  {author} {\bibinfo {author} {\bibfnamefont {R.}~\bibnamefont
  {Kusner}}, \bibinfo {author} {\bibfnamefont {W.}~\bibnamefont {Kusner}},
  \bibinfo {author} {\bibfnamefont {J.~C.}\ \bibnamefont {Lagarias}}, \ and\
  \bibinfo {author} {\bibfnamefont {S.}~\bibnamefont {Shlosman}},\ }\href
  {http://arxiv.org/abs/1611.10297} {\bibfield  {journal} {\bibinfo  {journal}
  {arXiv:1611.10297 [math.MG]}\ } (\bibinfo {year} {2016})},\ \Eprint
  {http://arxiv.org/abs/1611.10297} {arXiv:1611.10297 [cond-mat]} \BibitemShut
  {NoStop}%
\bibitem [{\citenamefont {Wales}\ and\ \citenamefont
  {Ulker}(2006)}]{Wales_Ulker_2006}%
  \BibitemOpen
  \bibfield  {author} {\bibinfo {author} {\bibfnamefont {D.~J.}\ \bibnamefont
  {Wales}}\ and\ \bibinfo {author} {\bibfnamefont {S.}~\bibnamefont {Ulker}},\
  }\href {\doibase 10.1103/PhysRevB.74.212101} {\bibfield  {journal} {\bibinfo
  {journal} {Phys. Rev. B}\ }\textbf {\bibinfo {volume} {74}},\ \bibinfo
  {pages} {212101} (\bibinfo {year} {2006})}\BibitemShut {NoStop}%
\bibitem [{\citenamefont {Wales}\ \emph {et~al.}(2009)\citenamefont {Wales},
  \citenamefont {McKay},\ and\ \citenamefont
  {Altschuler}}]{Wales_McKay_Altschuler_2009}%
  \BibitemOpen
  \bibfield  {author} {\bibinfo {author} {\bibfnamefont {D.~J.}\ \bibnamefont
  {Wales}}, \bibinfo {author} {\bibfnamefont {H.}~\bibnamefont {McKay}}, \ and\
  \bibinfo {author} {\bibfnamefont {E.~L.}\ \bibnamefont {Altschuler}},\ }\href
  {\doibase 10.1103/PhysRevB.79.224115} {\bibfield  {journal} {\bibinfo
  {journal} {Phys. Rev. B}\ }\textbf {\bibinfo {volume} {79}},\ \bibinfo
  {pages} {224115} (\bibinfo {year} {2009})}\BibitemShut {NoStop}%
\bibitem [{\citenamefont {Levin}(2000)}]{Levin_2000}%
  \BibitemOpen
  \bibfield  {author} {\bibinfo {author} {\bibfnamefont {Y.}~\bibnamefont
  {Levin}},\ }\href {\doibase https://doi.org/10.1016/S0378-4371(00)00459-3}
  {\bibfield  {journal} {\bibinfo  {journal} {Physica A: Statistical Mechanics
  and its Applications}\ }\textbf {\bibinfo {volume} {287}},\ \bibinfo {pages}
  {100 } (\bibinfo {year} {2000})}\BibitemShut {NoStop}%
\bibitem [{\citenamefont {Pusey}\ \emph {et~al.}(2009)\citenamefont {Pusey},
  \citenamefont {Zaccarelli}, \citenamefont {Valeriani}, \citenamefont {Sanz},
  \citenamefont {Poon},\ and\ \citenamefont {Cates}}]{Pusey-2009}%
  \BibitemOpen
  \bibfield  {author} {\bibinfo {author} {\bibfnamefont {P.~N.}\ \bibnamefont
  {Pusey}}, \bibinfo {author} {\bibfnamefont {E.}~\bibnamefont {Zaccarelli}},
  \bibinfo {author} {\bibfnamefont {C.}~\bibnamefont {Valeriani}}, \bibinfo
  {author} {\bibfnamefont {E.}~\bibnamefont {Sanz}}, \bibinfo {author}
  {\bibfnamefont {W.~C.~K.}\ \bibnamefont {Poon}}, \ and\ \bibinfo {author}
  {\bibfnamefont {M.~E.}\ \bibnamefont {Cates}},\ }\href {\doibase
  10.1098/rsta.2009.0181} {\bibfield  {journal} {\bibinfo  {journal}
  {Philosophical Transactions of the Royal Society of London A: Mathematical,
  Physical and Engineering Sciences}\ }\textbf {\bibinfo {volume} {367}},\
  \bibinfo {pages} {4993} (\bibinfo {year} {2009})},\ \Eprint
  {http://arxiv.org/abs/http://rsta.royalsocietypublishing.org/content/367/1909/4993.full.pdf}
  {http://rsta.royalsocietypublishing.org/content/367/1909/4993.full.pdf}
  \BibitemShut {NoStop}%
\bibitem [{\citenamefont
  {Jones}(1924)}]{Jones_DeterminationMolecularFields_1924}%
  \BibitemOpen
  \bibfield  {author} {\bibinfo {author} {\bibfnamefont {J.~E.}\ \bibnamefont
  {Jones}},\ }\href {\doibase 10.1098/rspa.1924.0082} {\bibfield  {journal}
  {\bibinfo  {journal} {Proc. Royal Soc. London A: Math., Phys. Eng. Sci.}\
  }\textbf {\bibinfo {volume} {106}},\ \bibinfo {pages} {463} (\bibinfo {year}
  {1924})}\BibitemShut {NoStop}%
\bibitem [{\citenamefont {Lennard-Jones}(1931)}]{Lennard-Jones_Cohesion_1931}%
  \BibitemOpen
  \bibfield  {author} {\bibinfo {author} {\bibfnamefont {J.~E.}\ \bibnamefont
  {Lennard-Jones}},\ }\href {\doibase 10.1088/0959-5309/43/5/301} {\bibfield
  {journal} {\bibinfo  {journal} {Proc. Phys. Soc.}\ }\textbf {\bibinfo
  {volume} {43}},\ \bibinfo {pages} {461} (\bibinfo {year} {1931})}\BibitemShut
  {NoStop}%
\bibitem [{\citenamefont {Trombach}\ \emph {et~al.}(2018)\citenamefont
  {Trombach}, \citenamefont {Hoy}, \citenamefont {Wales},\ and\ \citenamefont
  {Schwerdtfeger}}]{Trombach_2018}%
  \BibitemOpen
  \bibfield  {author} {\bibinfo {author} {\bibfnamefont {L.}~\bibnamefont
  {Trombach}}, \bibinfo {author} {\bibfnamefont {R.~S.}\ \bibnamefont {Hoy}},
  \bibinfo {author} {\bibfnamefont {D.~J.}\ \bibnamefont {Wales}}, \ and\
  \bibinfo {author} {\bibfnamefont {P.}~\bibnamefont {Schwerdtfeger}},\ }\href
  {\doibase 10.1103/PhysRevE.97.043309} {\bibfield  {journal} {\bibinfo
  {journal} {Phys. Rev. E}\ }\textbf {\bibinfo {volume} {97}},\ \bibinfo
  {pages} {043309} (\bibinfo {year} {2018})}\BibitemShut {NoStop}%
\bibitem [{\citenamefont {Mackay}(1962)}]{Mackay-1962}%
  \BibitemOpen
  \bibfield  {author} {\bibinfo {author} {\bibfnamefont {A.~L.}\ \bibnamefont
  {Mackay}},\ }\href {\doibase 10/cfw8n3} {\bibfield  {journal} {\bibinfo
  {journal} {Acta Cryst.}\ }\textbf {\bibinfo {volume} {15}},\ \bibinfo {pages}
  {916} (\bibinfo {year} {1962})}\BibitemShut {NoStop}%
\bibitem [{\citenamefont {Hoare}\ and\ \citenamefont
  {Pal}(1975)}]{Hoare_Physicalclustermechanics_1975}%
  \BibitemOpen
  \bibfield  {author} {\bibinfo {author} {\bibfnamefont {M.~R.}\ \bibnamefont
  {Hoare}}\ and\ \bibinfo {author} {\bibfnamefont {P.}~\bibnamefont {Pal}},\
  }\href {\doibase 10/d5pz4j} {\bibfield  {journal} {\bibinfo  {journal}
  {Advances in Physics}\ }\textbf {\bibinfo {volume} {24}},\ \bibinfo {pages}
  {645} (\bibinfo {year} {1975})}\BibitemShut {NoStop}%
\bibitem [{\citenamefont {Klots}\ \emph {et~al.}(1990)\citenamefont {Klots},
  \citenamefont {Winter}, \citenamefont {Parks},\ and\ \citenamefont
  {Riley}}]{Klots90}%
  \BibitemOpen
  \bibfield  {author} {\bibinfo {author} {\bibfnamefont {T.~D.}\ \bibnamefont
  {Klots}}, \bibinfo {author} {\bibfnamefont {B.~J.}\ \bibnamefont {Winter}},
  \bibinfo {author} {\bibfnamefont {E.~K.}\ \bibnamefont {Parks}}, \ and\
  \bibinfo {author} {\bibfnamefont {S.~J.}\ \bibnamefont {Riley}},\ }\href@noop
  {} {\bibfield  {journal} {\bibinfo  {journal} {J. Chem. Phys.}\ }\textbf
  {\bibinfo {volume} {92}},\ \bibinfo {pages} {2110} (\bibinfo {year}
  {1990})}\BibitemShut {NoStop}%
\bibitem [{\citenamefont {Uppenbrink}\ and\ \citenamefont
  {Wales}(1991)}]{Uppenbrink-1991}%
  \BibitemOpen
  \bibfield  {author} {\bibinfo {author} {\bibfnamefont {J.}~\bibnamefont
  {Uppenbrink}}\ and\ \bibinfo {author} {\bibfnamefont {D.~J.}\ \bibnamefont
  {Wales}},\ }\href {\doibase 10/fwvxzj} {\bibfield  {journal} {\bibinfo
  {journal} {J. Chem. Soc. - Faraday Trans.}\ }\textbf {\bibinfo {volume}
  {87}},\ \bibinfo {pages} {215} (\bibinfo {year} {1991})}\BibitemShut
  {NoStop}%
\bibitem [{\citenamefont {van~de Waal}(1993)}]{vandewaal93}%
  \BibitemOpen
  \bibfield  {author} {\bibinfo {author} {\bibfnamefont {B.~W.}\ \bibnamefont
  {van~de Waal}},\ }\href@noop {} {\bibfield  {journal} {\bibinfo  {journal}
  {J. Chem. Phys.}\ }\textbf {\bibinfo {volume} {98}},\ \bibinfo {pages} {4909}
  (\bibinfo {year} {1993})}\BibitemShut {NoStop}%
\bibitem [{\citenamefont {Wales}\ \emph {et~al.}(1996)\citenamefont {Wales},
  \citenamefont {Munro},\ and\ \citenamefont {Doye}}]{Wales-1996a2}%
  \BibitemOpen
  \bibfield  {author} {\bibinfo {author} {\bibfnamefont {D.~J.}\ \bibnamefont
  {Wales}}, \bibinfo {author} {\bibfnamefont {L.~J.}\ \bibnamefont {Munro}}, \
  and\ \bibinfo {author} {\bibfnamefont {J.~P.~K.}\ \bibnamefont {Doye}},\
  }\href {\doibase 10.1039/DT9960000611} {\bibfield  {journal} {\bibinfo
  {journal} {J. Chem. Soc.{,} Dalton Trans.}\ ,\ \bibinfo {pages} {611}}
  (\bibinfo {year} {1996})}\BibitemShut {NoStop}%
\bibitem [{\citenamefont {Wales}\ and\ \citenamefont
  {Munro}(1996)}]{wales_1996a3}%
  \BibitemOpen
  \bibfield  {author} {\bibinfo {author} {\bibfnamefont {D.~J.}\ \bibnamefont
  {Wales}}\ and\ \bibinfo {author} {\bibfnamefont {L.~J.}\ \bibnamefont
  {Munro}},\ }\href@noop {} {\bibfield  {journal} {\bibinfo  {journal} {J.
  Chem. Phys.}\ }\textbf {\bibinfo {volume} {100}},\ \bibinfo {pages} {2053}
  (\bibinfo {year} {1996})}\BibitemShut {NoStop}%
\bibitem [{\citenamefont {Baxter}(1968)}]{baxter68}%
  \BibitemOpen
  \bibfield  {author} {\bibinfo {author} {\bibfnamefont {R.~J.}\ \bibnamefont
  {Baxter}},\ }\href {\doibase 10.1063/1.1670482} {\bibfield  {journal}
  {\bibinfo  {journal} {J. Chem. Phys.}\ }\textbf {\bibinfo {volume} {49}},\
  \bibinfo {pages} {2770} (\bibinfo {year} {1968})}\BibitemShut {NoStop}%
\bibitem [{\citenamefont {Gazzillo}\ and\ \citenamefont
  {Giacometti}(2004)}]{Gazzillo_2004}%
  \BibitemOpen
  \bibfield  {author} {\bibinfo {author} {\bibfnamefont {D.}~\bibnamefont
  {Gazzillo}}\ and\ \bibinfo {author} {\bibfnamefont {A.}~\bibnamefont
  {Giacometti}},\ }\href {\doibase 10.1063/1.1645781} {\bibfield  {journal}
  {\bibinfo  {journal} {J. Chem. Phys.}\ }\textbf {\bibinfo {volume} {120}},\
  \bibinfo {pages} {4742} (\bibinfo {year} {2004})},\ \Eprint
  {http://arxiv.org/abs/https://doi.org/10.1063/1.1645781}
  {https://doi.org/10.1063/1.1645781} \BibitemShut {NoStop}%
\bibitem [{\citenamefont {Stell}(1991)}]{Stell_1991}%
  \BibitemOpen
  \bibfield  {author} {\bibinfo {author} {\bibfnamefont {G.}~\bibnamefont
  {Stell}},\ }\href {\doibase 10.1007/BF01030007} {\bibfield  {journal}
  {\bibinfo  {journal} {J. Stat. Phys.}\ }\textbf {\bibinfo {volume} {63}},\
  \bibinfo {pages} {1203} (\bibinfo {year} {1991})}\BibitemShut {NoStop}%
\bibitem [{\citenamefont {Jamnik}(1996)}]{Jamnik_1996}%
  \BibitemOpen
  \bibfield  {author} {\bibinfo {author} {\bibfnamefont {A.}~\bibnamefont
  {Jamnik}},\ }\href {\doibase 10.1063/1.472940} {\bibfield  {journal}
  {\bibinfo  {journal} {J. Chem. Phys.}\ }\textbf {\bibinfo {volume} {105}},\
  \bibinfo {pages} {10511} (\bibinfo {year} {1996})},\ \Eprint
  {http://arxiv.org/abs/https://doi.org/10.1063/1.472940}
  {https://doi.org/10.1063/1.472940} \BibitemShut {NoStop}%
\bibitem [{\citenamefont {Santos}\ \emph {et~al.}(1998)\citenamefont {Santos},
  \citenamefont {Yuste},\ and\ \citenamefont {de~Haro}}]{Santos_1998}%
  \BibitemOpen
  \bibfield  {author} {\bibinfo {author} {\bibfnamefont {A.}~\bibnamefont
  {Santos}}, \bibinfo {author} {\bibfnamefont {S.~B.}\ \bibnamefont {Yuste}}, \
  and\ \bibinfo {author} {\bibfnamefont {M.~L.}\ \bibnamefont {de~Haro}},\
  }\href {\doibase 10.1063/1.477328} {\bibfield  {journal} {\bibinfo  {journal}
  {J. Chem. Phys.}\ }\textbf {\bibinfo {volume} {109}},\ \bibinfo {pages}
  {6814} (\bibinfo {year} {1998})},\ \Eprint
  {http://arxiv.org/abs/https://doi.org/10.1063/1.477328}
  {https://doi.org/10.1063/1.477328} \BibitemShut {NoStop}%
\bibitem [{\citenamefont {Hoy}\ and\ \citenamefont
  {O'Hern}(2010)}]{Hoy_MinimalEnergyPackings_2010}%
  \BibitemOpen
  \bibfield  {author} {\bibinfo {author} {\bibfnamefont {R.~S.}\ \bibnamefont
  {Hoy}}\ and\ \bibinfo {author} {\bibfnamefont {C.~S.}\ \bibnamefont
  {O'Hern}},\ }\href {\doibase 10.1103/PhysRevLett.105.068001} {\bibfield
  {journal} {\bibinfo  {journal} {Phys. Rev. Lett.}\ }\textbf {\bibinfo
  {volume} {105}},\ \bibinfo {pages} {068001} (\bibinfo {year}
  {2010})}\BibitemShut {NoStop}%
\bibitem [{\citenamefont {Arkus}\ \emph {et~al.}(2009)\citenamefont {Arkus},
  \citenamefont {Manoharan},\ and\ \citenamefont
  {Brenner}}]{Arkus_Minimalenergyclusters_2009}%
  \BibitemOpen
  \bibfield  {author} {\bibinfo {author} {\bibfnamefont {N.}~\bibnamefont
  {Arkus}}, \bibinfo {author} {\bibfnamefont {V.~N.}\ \bibnamefont
  {Manoharan}}, \ and\ \bibinfo {author} {\bibfnamefont {M.~P.}\ \bibnamefont
  {Brenner}},\ }\href {\doibase 10.1103/PhysRevLett.103.118303} {\bibfield
  {journal} {\bibinfo  {journal} {Phys. Rev. Lett.}\ }\textbf {\bibinfo
  {volume} {103}},\ \bibinfo {pages} {118303} (\bibinfo {year}
  {2009})}\BibitemShut {NoStop}%
\bibitem [{\citenamefont {Meng}\ \emph {et~al.}(2010)\citenamefont {Meng},
  \citenamefont {Arkus}, \citenamefont {Brenner},\ and\ \citenamefont
  {Manoharan}}]{Arkus-2010}%
  \BibitemOpen
  \bibfield  {author} {\bibinfo {author} {\bibfnamefont {G.}~\bibnamefont
  {Meng}}, \bibinfo {author} {\bibfnamefont {N.}~\bibnamefont {Arkus}},
  \bibinfo {author} {\bibfnamefont {M.~P.}\ \bibnamefont {Brenner}}, \ and\
  \bibinfo {author} {\bibfnamefont {V.~N.}\ \bibnamefont {Manoharan}},\ }\href
  {\doibase 10/b5c3d5} {\bibfield  {journal} {\bibinfo  {journal} {Science}\
  }\textbf {\bibinfo {volume} {327}},\ \bibinfo {pages} {560} (\bibinfo {year}
  {2010})}\BibitemShut {NoStop}%
\bibitem [{\citenamefont {Arkus}\ \emph {et~al.}(2011)\citenamefont {Arkus},
  \citenamefont {Manoharan},\ and\ \citenamefont
  {Brenner}}]{Arkus_DerivingFiniteSphere_2011}%
  \BibitemOpen
  \bibfield  {author} {\bibinfo {author} {\bibfnamefont {N.}~\bibnamefont
  {Arkus}}, \bibinfo {author} {\bibfnamefont {V.}~\bibnamefont {Manoharan}}, \
  and\ \bibinfo {author} {\bibfnamefont {M.}~\bibnamefont {Brenner}},\ }\href
  {\doibase 10.1137/100784424} {\bibfield  {journal} {\bibinfo  {journal} {SIAM
  J. Discrete Math.}\ }\textbf {\bibinfo {volume} {25}},\ \bibinfo {pages}
  {1860} (\bibinfo {year} {2011})}\BibitemShut {NoStop}%
\bibitem [{\citenamefont {Hoy}\ \emph {et~al.}(2012)\citenamefont {Hoy},
  \citenamefont {Harwayne-Gidansky},\ and\ \citenamefont
  {O'Hern}}]{Hoy_Structurefinitesphere_2012}%
  \BibitemOpen
  \bibfield  {author} {\bibinfo {author} {\bibfnamefont {R.~S.}\ \bibnamefont
  {Hoy}}, \bibinfo {author} {\bibfnamefont {J.}~\bibnamefont
  {Harwayne-Gidansky}}, \ and\ \bibinfo {author} {\bibfnamefont {C.~S.}\
  \bibnamefont {O'Hern}},\ }\href {\doibase 10.1103/PhysRevE.85.051403}
  {\bibfield  {journal} {\bibinfo  {journal} {Phys. Rev. E}\ }\textbf {\bibinfo
  {volume} {85}},\ \bibinfo {pages} {051403} (\bibinfo {year}
  {2012})}\BibitemShut {NoStop}%
\bibitem [{\citenamefont {Hayes}(2012)}]{Hayes_ScienceStickySpheres_2012}%
  \BibitemOpen
  \bibfield  {author} {\bibinfo {author} {\bibfnamefont {B.}~\bibnamefont
  {Hayes}},\ }\href {\doibase 10.1511/2012.99.442} {\bibfield  {journal}
  {\bibinfo  {journal} {Am. Scientist}\ }\textbf {\bibinfo {volume} {100}},\
  \bibinfo {pages} {442} (\bibinfo {year} {2012})}\BibitemShut {NoStop}%
\bibitem [{\citenamefont {Holmes-Cerfon}\ \emph {et~al.}(2013)\citenamefont
  {Holmes-Cerfon}, \citenamefont {Gortler},\ and\ \citenamefont
  {Brenner}}]{Holmes-Cerfon_geometricalapproachcomputing_2013}%
  \BibitemOpen
  \bibfield  {author} {\bibinfo {author} {\bibfnamefont {M.}~\bibnamefont
  {Holmes-Cerfon}}, \bibinfo {author} {\bibfnamefont {S.~J.}\ \bibnamefont
  {Gortler}}, \ and\ \bibinfo {author} {\bibfnamefont {M.~P.}\ \bibnamefont
  {Brenner}},\ }\href {\doibase 10.1073/pnas.1211720110} {\bibfield  {journal}
  {\bibinfo  {journal} {Proc. Natl. Acad. Sci. U.S.A.}\ }\textbf {\bibinfo
  {volume} {110}},\ \bibinfo {pages} {E5} (\bibinfo {year} {2013})},\ \Eprint
  {http://arxiv.org/abs/23248296} {23248296} \BibitemShut {NoStop}%
\bibitem [{\citenamefont
  {Holmes-Cerfon}(2016)}]{Holmes-Cerfon_EnumeratingRigidSphere_2016}%
  \BibitemOpen
  \bibfield  {author} {\bibinfo {author} {\bibfnamefont {M.}~\bibnamefont
  {Holmes-Cerfon}},\ }\href {\doibase 10.1137/140982337} {\bibfield  {journal}
  {\bibinfo  {journal} {SIAM Review}\ }\textbf {\bibinfo {volume} {58}},\
  \bibinfo {pages} {229} (\bibinfo {year} {2016})}\BibitemShut {NoStop}%
\bibitem [{\citenamefont
  {Holmes-Cerfon}(2017)}]{Holmes-Cerfon_StickySphereClusters_2017}%
  \BibitemOpen
  \bibfield  {author} {\bibinfo {author} {\bibfnamefont {M.}~\bibnamefont
  {Holmes-Cerfon}},\ }\href {\doibase 10.1146/annurev-conmatphys-031016-025357}
  {\bibfield  {journal} {\bibinfo  {journal} {Ann. Rev. Cond. Matter Phys.}\
  }\textbf {\bibinfo {volume} {8}},\ \bibinfo {pages} {77} (\bibinfo {year}
  {2017})}\BibitemShut {NoStop}%
\bibitem [{\citenamefont {Kallus}\ and\ \citenamefont
  {Holmes-Cerfon}(2017)}]{Kallus_Freeenergysingular_2017}%
  \BibitemOpen
  \bibfield  {author} {\bibinfo {author} {\bibfnamefont {Y.}~\bibnamefont
  {Kallus}}\ and\ \bibinfo {author} {\bibfnamefont {M.}~\bibnamefont
  {Holmes-Cerfon}},\ }\href {\doibase 10.1103/PhysRevE.95.022130} {\bibfield
  {journal} {\bibinfo  {journal} {Physical Review E}\ }\textbf {\bibinfo
  {volume} {95}},\ \bibinfo {pages} {022130} (\bibinfo {year}
  {2017})}\BibitemShut {NoStop}%
\bibitem [{\citenamefont {Hoy}(2015)}]{Hoy_Structuredynamicsmodel_2015}%
  \BibitemOpen
  \bibfield  {author} {\bibinfo {author} {\bibfnamefont {R.~S.}\ \bibnamefont
  {Hoy}},\ }\href {\doibase 10.1103/PhysRevE.91.012303} {\bibfield  {journal}
  {\bibinfo  {journal} {Physical Review E}\ }\textbf {\bibinfo {volume} {91}},\
  \bibinfo {pages} {012303} (\bibinfo {year} {2015})}\BibitemShut {NoStop}%
\bibitem [{\citenamefont {Wales}(2001)}]{Wales_MicroscopicBasisGlobal_2001}%
  \BibitemOpen
  \bibfield  {author} {\bibinfo {author} {\bibfnamefont {D.~J.}\ \bibnamefont
  {Wales}},\ }\href {\doibase 10.1126/science.1062565} {\bibfield  {journal}
  {\bibinfo  {journal} {Science}\ }\textbf {\bibinfo {volume} {293}},\ \bibinfo
  {pages} {2067} (\bibinfo {year} {2001})}\BibitemShut {NoStop}%
\bibitem [{\citenamefont {Cordella}\ \emph {et~al.}(2004)\citenamefont
  {Cordella}, \citenamefont {Foggia}, \citenamefont {Sansone},\ and\
  \citenamefont {Vento}}]{Cordella_SubGraphIsomorphism_2004}%
  \BibitemOpen
  \bibfield  {author} {\bibinfo {author} {\bibfnamefont {L.~P.}\ \bibnamefont
  {Cordella}}, \bibinfo {author} {\bibfnamefont {P.}~\bibnamefont {Foggia}},
  \bibinfo {author} {\bibfnamefont {C.}~\bibnamefont {Sansone}}, \ and\
  \bibinfo {author} {\bibfnamefont {M.}~\bibnamefont {Vento}},\ }\href
  {\doibase 10.1109/TPAMI.2004.75} {\bibfield  {journal} {\bibinfo  {journal}
  {IEEE Transactions on Pattern Analysis and Machine Intelligence}\ }\textbf
  {\bibinfo {volume} {26}},\ \bibinfo {pages} {1367} (\bibinfo {year}
  {2004})}\BibitemShut {NoStop}%
\bibitem [{\citenamefont {Siek}\ \emph {et~al.}(2002)\citenamefont {Siek},
  \citenamefont {Lee},\ and\ \citenamefont {Lumsdaine}}]{_boost_2002}%
  \BibitemOpen
  \bibfield  {author} {\bibinfo {author} {\bibfnamefont {J.}~\bibnamefont
  {Siek}}, \bibinfo {author} {\bibfnamefont {L.-Q.}\ \bibnamefont {Lee}}, \
  and\ \bibinfo {author} {\bibfnamefont {A.}~\bibnamefont {Lumsdaine}},\
  }\href@noop {} {\emph {\bibinfo {title} {The {{Boost Graph Library}}: User
  Guide and Reference Manual}}},\ !!\ (\bibinfo  {publisher} {{Addison-Wesley
  Longman Publishing Co., Inc.}},\ \bibinfo {address} {Boston, MA, USA},\
  \bibinfo {year} {2002})\BibitemShut {NoStop}%
\bibitem [{\citenamefont {King}(2009)}]{King_DlibmlMachineLearning_2009}%
  \BibitemOpen
  \bibfield  {author} {\bibinfo {author} {\bibfnamefont {D.~E.}\ \bibnamefont
  {King}},\ }\href@noop {} {\bibfield  {journal} {\bibinfo  {journal} {Journal
  of Machine Learning Research}\ }\textbf {\bibinfo {volume} {10}},\ \bibinfo
  {pages} {1755} (\bibinfo {year} {2009})}\BibitemShut {NoStop}%
\bibitem [{\citenamefont {Sch{\"u}tte}\ and\ \citenamefont {van~der
  Waerden}(1951)}]{Schutte_1951}%
  \BibitemOpen
  \bibfield  {author} {\bibinfo {author} {\bibfnamefont {K.}~\bibnamefont
  {Sch{\"u}tte}}\ and\ \bibinfo {author} {\bibfnamefont {B.~L.}\ \bibnamefont
  {van~der Waerden}},\ }\href {\doibase 10.1007/BF02054944} {\bibfield
  {journal} {\bibinfo  {journal} {Math. Ann.}\ }\textbf {\bibinfo {volume}
  {123}},\ \bibinfo {pages} {96} (\bibinfo {year} {1951})}\BibitemShut
  {NoStop}%
\bibitem [{\citenamefont {Steinitz}(1922)}]{Steinitz1922}%
  \BibitemOpen
  \bibfield  {author} {\bibinfo {author} {\bibfnamefont {E.}~\bibnamefont
  {Steinitz}},\ }\href@noop {} {\bibfield  {journal} {\bibinfo  {journal}
  {Enzyklop{\"a}die der mathematischen Wissenschaften, Band 3 (Geometries)}\
  }\textbf {\bibinfo {volume} {3}},\ \bibinfo {pages} {1} (\bibinfo {year}
  {1922})}\BibitemShut {NoStop}%
\bibitem [{\citenamefont {Doye}\ \emph {et~al.}(1995)\citenamefont {Doye},
  \citenamefont {Wales},\ and\ \citenamefont {Berry}}]{Doye_1995}%
  \BibitemOpen
  \bibfield  {author} {\bibinfo {author} {\bibfnamefont {J.~P.~K.}\
  \bibnamefont {Doye}}, \bibinfo {author} {\bibfnamefont {D.~J.}\ \bibnamefont
  {Wales}}, \ and\ \bibinfo {author} {\bibfnamefont {R.~S.}\ \bibnamefont
  {Berry}},\ }\href {\doibase 10.1063/1.470729} {\bibfield  {journal} {\bibinfo
   {journal} {J. Chem. Phys.}\ }\textbf {\bibinfo {volume} {103}},\ \bibinfo
  {pages} {4234} (\bibinfo {year} {1995})},\ \Eprint
  {http://arxiv.org/abs/https://doi.org/10.1063/1.470729}
  {https://doi.org/10.1063/1.470729} \BibitemShut {NoStop}%
\bibitem [{\citenamefont {Wales}\ and\ \citenamefont
  {Doye}(1996)}]{Wales_1996a1}%
  \BibitemOpen
  \bibfield  {author} {\bibinfo {author} {\bibfnamefont {D.~J.}\ \bibnamefont
  {Wales}}\ and\ \bibinfo {author} {\bibfnamefont {J.~P.}\ \bibnamefont
  {Doye}},\ }in\ \href@noop {} {\emph {\bibinfo {booktitle} {Large Clusters of
  Atoms and Molecules}}}\ (\bibinfo  {publisher} {Springer},\ \bibinfo {year}
  {1996})\ pp.\ \bibinfo {pages} {241--279}\BibitemShut {NoStop}%
\bibitem [{\citenamefont {Doye}\ and\ \citenamefont
  {Wales}(1996)}]{Doye_1996a}%
  \BibitemOpen
  \bibfield  {author} {\bibinfo {author} {\bibfnamefont {J.~P.~K.}\
  \bibnamefont {Doye}}\ and\ \bibinfo {author} {\bibfnamefont {D.~J.}\
  \bibnamefont {Wales}},\ }\href
  {http://stacks.iop.org/0953-4075/29/i=21/a=002} {\bibfield  {journal}
  {\bibinfo  {journal} {J. Phys. B: At. Mol. Opt. Phys.}\ }\textbf {\bibinfo
  {volume} {29}},\ \bibinfo {pages} {4859} (\bibinfo {year}
  {1996})}\BibitemShut {NoStop}%
\bibitem [{\citenamefont {P.~K.~Doye}\ and\ \citenamefont
  {J.~Wales}(1997)}]{Doye_1997a}%
  \BibitemOpen
  \bibfield  {author} {\bibinfo {author} {\bibfnamefont {J.}~\bibnamefont
  {P.~K.~Doye}}\ and\ \bibinfo {author} {\bibfnamefont {D.}~\bibnamefont
  {J.~Wales}},\ }\href {\doibase 10.1039/A706221D} {\bibfield  {journal}
  {\bibinfo  {journal} {J. Chem. Soc.{,} Faraday Trans.}\ }\textbf {\bibinfo
  {volume} {93}},\ \bibinfo {pages} {4233} (\bibinfo {year}
  {1997})}\BibitemShut {NoStop}%
\bibitem [{\citenamefont {de~Heer}(1993)}]{Heer-1993}%
  \BibitemOpen
  \bibfield  {author} {\bibinfo {author} {\bibfnamefont {W.~A.}\ \bibnamefont
  {de~Heer}},\ }\href {\doibase 10.1103/RevModPhys.65.611} {\bibfield
  {journal} {\bibinfo  {journal} {Rev. Mod. Phys.}\ }\textbf {\bibinfo {volume}
  {65}},\ \bibinfo {pages} {611} (\bibinfo {year} {1993})}\BibitemShut
  {NoStop}%
\end{thebibliography}%

\end{document}